\documentclass[reprint,prd,onecolumn,notitlepage,11pt,nofootinbib]{revtex4-1}
\usepackage[english]{babel}
\usepackage{amsmath,amssymb,graphicx,bm,lipsum}
\usepackage[colorlinks=true,citecolor=blue,linkcolor=blue,urlcolor=blue]{hyperref} 
\usepackage[font={small},flushleft,indent]{caption}
\usepackage{setspace}

\begin{document}

\title{Ultralight scalar dark matter versus non-adiabatic perfect fluid dark matter in pulsar timing}

\author{Qing-Hua Zhu}
\email{zhuqh@cqu.edu.cn} 
\affiliation{School of Physics, Chongqing University, Chongqing 401331, China} 
 
\begin{abstract} 
Recent evidence for stochastic gravitational waves reported by pulsar timing array (PTA) collaborations might open a new window for studying cosmology and astrophysical phenomena. In addition to signals from gravitational waves, there is motivation to explore residual signals from oscillating dark matter, which might partially comprise the galactic halo. We investigate fluctuations in pulsar timing originating from the coherent oscillation of scalar dark matter up to the subleading-order correction of $\mathcal{O}(k/m)$, as well as from acoustic oscillations of non-adiabatic perfect fluid dark matter. 
Both types of dark matter can induce the Newtonian potential and curvature perturbations, thereby affecting pulsar timing. We show distinctive signatures in pulsar timing residuals and angular correlations in the PTA frequency band and considering the known distances of pulsars.
For scalar dark matter, both the timing residuals and the angular correlation are sensitive to small variations in the distance, $\delta L$, due to the subleading-order correction of $\mathcal{O}(k/m)$. In contrast, for perfect fluid dark matter, it is insensitive to the $\delta L$. For deterministic sources from scalar dark matter, the distance of a pulsar has influence on the degree of directional dependence of timing residuals, significantly. 
For stochastic sources from perfect fluid dark matter, the angular correlation tends to a constant and enhances only when the pulsar pair is very close to each other. In this sense, perfect fluid dark matter is shown to be a more suitable physical origin for monopolar signals in angular correlations compared to the scalar dark matter.

\end{abstract} 

\maketitle

\section{introduction}

Recent evidence for stochastic gravitational waves in nHz frequency band might open a new window for studying cosmology in the early universe and astrophysical phenomena on sub-galactic scales \cite{Xu:2023wog,Zic:2023gta,EuropeanPulsarTimingArray:2023egv,EPTA:2023akd,EPTA:2023fyk,EPTA:2023gyr,EPTA:2023xxk,Reardon:2023gzh,NANOGrav:2023gor,NANOGrav:2023hfp,NANOGrav:2023hvm,NANOGrav:2023tcn,NANOGrav:2023pdq}. A notable aspect of pulsar timing array (PTA) observations is that the pulsar timing has the capacity to reflect metric perturbations via spatially correlated signals, i.e., the Hellings-Downs (HD) curve \cite{Hellings:1983fr}. This capability enables the distinction of gravitational waves from other types of fluctuations \cite{2008ApJ...685.1304L}, including those originating from dark matter \cite{Khmelnitsky:2013lxt,EPTA:2023xxk,NANOGrav:2023hvm}.

Dark matter plays a crucial role in modern cosmology and astrophysical physics.  The standard cosmological model, known as the Lambda cold dark matter ($\Lambda$CDM) model, has been well-tested through the observations such as the cosmic microwave background \cite{Planck:2018vyg} and large-scale structure \cite{WMAP:2012nax}, whereas it appears to be inconsistent with observations on sub-galactic scales \cite{Weinberg:2013aya}. 
The ultralight scalar dark matter can be a cold dark matter candidate, exhibiting very distinct behavior on small scales, which might resolve discrepancies in small-scale observations \cite{Park:2012ru,Hwang:2009js,Pontzen:2011ty,Khmelnitsky:2013lxt,Hlozek:2014lca,Pontzen:2014lma,Marsh:2015xka,Nadler:2023nrd}. A decade ago, it was found that pulsar timing could be influenced by the coherent oscillation of ultralight scalar field at a frequency proportional to its mass \cite{Khmelnitsky:2013lxt}. Consequently, it was expected that dark matter signals could be detected based on PTA data sets \cite{Porayko:2014rfa,Porayko:2018sfa,Kato:2019bqz,Kaplan:2022lmz}. Recent studies have been further extended to ultralight vector dark matter \cite{Nomura:2019cvc,PPTA:2021uzb,PPTA:2022eul,Sun:2021yra,Unal:2022ooa} and ultralight tensor dark matter \cite{Armaleo:2020yml,Sun:2021yra,Unal:2022ooa,Wu:2023dnp}. The detectability of ultralight dark matter has been elucidated in pioneer studies. However, there has been rare investigation into similar mechanisms for CDM (formulated by a perfect fluid with $w=0$). Hence, the primary motivation of this paper is to explore the potential oscillatory features of CDM that might have an effect on pulsar timing.

It is noteworthy that there are two competitive mechanisms regarding how ultralight dark matter affects pulsar timing \cite{NANOGrav:2023hvm}: i) interaction between dark matter and normal matter \cite{Graham:2015ifn,Armaleo:2020yml,Kaplan:2022lmz,PPTA:2021uzb,Sun:2021yra,Unal:2022ooa}, and ii) pure gravitational effect \cite{Khmelnitsky:2013lxt,Aoki:2016mtn,Porayko:2018sfa,Kato:2019bqz,Nomura:2019cvc,PPTA:2022eul,Sun:2021yra,Unal:2022ooa,Wu:2023dnp}. In the latter, the Einstein field equation, sourced by the dark matter, should be perturbed to the second order. The dynamical equations indicate that the oscillating ultralight dark matter can generate metric fluctuations.   
In this study, we investigate the pulsar timing residuals influenced by scalar dark matter and non-adiabatic perfect fluid dark matter within the framework of mechanism ii), by employing a more rigorous formalism \cite{Mukhanov:1990me,Malik:2008im,Zhu:2022bwf}. For scalar dark matter, this formalism enables us to extensively calculate timing residuals to the subleading-order correction of $\mathcal{O}(k/m)$. And it is found that the timing residual turns out to be directional-dependent, due to the subleading terms. For perfect fluid dark matter, we derive the equation of state parameter $w=0$ and set the effective sound speed $c_{\rm s} \neq 0$, in order to let it serve as a type of CDM under non-adiabatic conditions \cite{Arena:2006nn,Buldgen_2015,Hu:2015dea}. 
It can be an effective fluid description of ultralight axion when the observational timescale is much longer than the oscillation periods \cite{Hwang:2009js,Park:2012ru,Marsh:2015xka,Passaglia:2022bcr}.
In this scenario, it is found that acoustic oscillations of perfect fluid dark matter should occur within our galaxy. Consequently, the oscillating dark matter can generate metric perturbations, influencing pulsar timing.


Besides above deterministic sources, the spatially correlated signals in pulsar timing can be used to detect stochastic sources \cite{Armaleo:2020yml}. Specifically, by spatially averaging all pairs of pulsars at a given angular separation, the angular correlation formulated by the overlap reduction function (ORF) can be derived \cite{Allen:2022dzg}. Using this approach, the angular correlations originating from ultralight dark matters have also been referred to as deformed Hellings-Downs curves \cite{Omiya:2023bio,Cai:2024thd}. Alternatively, one can also employ the ensemble-average approach by considering the dark matter field as a stochastic variable \cite{Luu:2023rgg,Kim:2023kyy}. Above two approaches would yield consistent angular correlations \cite{Allen:2022dzg}. In this study, we adopt the ensemble-average approach to study the angular correlations originating from the coherent oscillation of scalar dark matter to the subleading-order correction of $\mathcal{O}(k/m)$, as well as from the acoustic oscillation of non-adiabatic perfect fluid dark matter. It is found that the angular correlation originating from scalar dark matter is sensitive to distance of the pulsars, while the that for perfect fluid dark matter is not. It is expected that the distinctive signatures could be used to identify different types of dark matter in the future.  

The rest of the paper is organized as follows. In Sec.~\ref{II}, we present the dynamical evolution of metric perturbations induced by scalar dark matter and perfect fluid dark matter, based on perturbing Einstein field equation to the second order. In Sec.~\ref{III}, we calculate the timing residuals and angular correlations originating from scalar dark matter and perfect fluid dark matter. In Sec.~\ref{IV}, we compare the these dark matter models in PTA observations. Finally, in Sec.~\ref{V}, conclusions and discussions are summarized.

\
  
\section{Metric perturbations induced by oscillating dark matter \label{II}}

In this section, we derive and solve the equations of motion for metric perturbations originating from the coherent oscillation of scalar dark matter, as well as the acoustic oscillation of perfect fluid dark matter. For scalar dark matter, pioneering work obatined results in the limit of $k/m \ll 1$, where $m$ is the particle mass and $k (\equiv \lambda_\text{dB}^{-1})$ is the characteristic momentum derived from the de Broglie wavelength $\lambda_\text{dB}$ \cite{Khmelnitsky:2013lxt}. Since $k/m \simeq v_\text{Gal}$, where the typical dark matter velocity in our galaxy $v_\text{Gal} (\approx 10^{-3}c)$ is not very small, we extend the pioneering study to the subleading order of $k/m$.

The perturbed Minkowski metric to the second order in the Newtonian gauge is given by
\begin{eqnarray}
\mathrm{d} s^2 & = & - (1 + 2 \phi^{(1)} + \phi^{(2)}) \mathrm{d} t^2 + \left(
\delta_{i j} (1 - 2 \psi^{(1)} - \psi^{(2)}) + \frac{1}{2} h_{i
j} \right) \mathrm{d} x^i \mathrm{d} x^j + V_i \mathrm{d} t \mathrm{d} x^i ~, \label{met}
\end{eqnarray}
where $V_i (\equiv V_i^{(2)})$ is the second-order vector perturbation, $h_{i
j} (\equiv h_{i j}^{(2)})$ is the second-order tensor
perturbation, and $\phi^{(n)}$ and $\psi^{(n)}$ are the $n$th-order Newtonian
potential perturbation and curvature perturbation, respectively. Because the dark matter would not generate first-order vector or tensor perturbations, one can obtain $V^{(1)}_i = h_{i j}^{(1)} = 0$.

In subsequent parts, we will present evolution of the metric perturbations sourced by scalar dark matter and perfect fluid dark matter.

\subsection{Ultralight scalar dark matter}

The energy-momentum tensor of the scalar field $\varphi$ is given by
$
T_{\mu \nu}^{({\rm sdm})} = \nabla_{\mu} \varphi \nabla_{\nu} \varphi -
\frac{1}{2} g_{\mu \nu} (m^2 \varphi^2 + \nabla_{\lambda} \varphi
\nabla^{\lambda} \varphi) ~,
$
where $m$ is the scalar field mass, and $\nabla$ is the covariant derivative. The scalar field can be expanded in the form of $\varphi = \varphi^{(1)} + (1 /2) \varphi^{(2)} +\mathcal{O} (3)$. One can obtain $\varphi^{(0)} = 0$ due to the Minkowski background. Expanding the equation of energy-momentum tensor conservation $\nabla_{\mu} T^{\mu \nu} = 0$ to the second order, one can obtain the equation of motion for the scalar field, namely,
\begin{eqnarray}
  {\varphi^{(1)}}'' + (m^2 - \Delta) \varphi^{(1)} & = & 0 ~,
  \label{DTsdm1}
\end{eqnarray}
where $\Delta$ denotes the Laplace operator, and $'$ denotes the time derivative. The above equation of motion suggests that the scalar field oscillates in both space and time. By employing the expansion of $T_{\mu \nu}^{({\rm sdm})}$, one can derive the first-order Einstein field equation, which yields a trivial solution $\psi^{ (1)} = \phi^{(1)} = 0$. Evaluating the second-order Einstein field equations in space-space components, we obtain
\begin{subequations}
  \begin{eqnarray}
    \frac{1}{4} (h''_{ab} - \Delta h_{ab}) & = & \kappa
    \Lambda_{ab}^{c   d} S^{({\rm sdm})}_{c   d} ~, 
    \label{moesdm1}\\
    - \frac{1}{4} V_a' & = & \kappa \Delta^{- 1} (\delta_a^c - \partial_a
    \Delta^{- 1} \partial^c) \partial^d S^{({\rm sdm})}_{c   d} ~, \\
    {\psi^{(2)}}'' & = & \kappa \partial^c \Delta^{- 1} \partial^d
    S^{({\rm sdm})}_{c   d} ~, \\
    \frac{1}{2} (\psi^{(2)} - \phi^{(2)}) & = & \frac{\kappa}{2} \Delta^{- 1} (3
    \partial^c \Delta^{- 1} \partial^d - \delta^{c   d}) S_{c  
    d}^{({\rm sdm})} ~,  \label{moesdm4}
  \end{eqnarray} \label{moesdm}
\end{subequations}
where we have employed the helicity decomposition, $\Lambda_{ab}^{c
d}$ is the transverse-traceless operator, $\kappa$ is the Einstein gravitational constant, and
\begin{eqnarray}
S^{({\rm sdm})}_{b c} & = & \partial_b \varphi^{(1)} \partial_c
\varphi^{(1)} - \frac{1}{2} \delta_{b c} \left( m^2
(\varphi^{(1)})^2 - \left( {\varphi^{(1)}}' \right)^2 + \partial_d
\varphi^{(1)} \partial^d \varphi^{(1)} \right) ~. \label{Ssdm}
\end{eqnarray}
It shows that Eqs.~(\ref{moesdm}) do not involve the second-order scalar field $\varphi^{(2)}$. And the scalar, vector, and tensor metric perturbations are all induced by the scalar field $\varphi^{(1)}$. 

The solution of Eq.~(\ref{DTsdm1}) can be expressed in the form of plane wave expansion, namely
\begin{eqnarray}
  \varphi^{(1)} & = & \int \frac{ {\rm d}^3 k}{(2 \pi)^3} \left\{
  \bar{\varphi}_{\textbf{k}} e^{- i   \left( w_k t - \textbf{k} \cdot
  \textbf{x} \right)} + \bar{\varphi}_{- \textbf{k}} e^{i   \left( w_k
  t - \textbf{k} \cdot \textbf{x} \right)} \right\} ~, \label{pwescalar}
\end{eqnarray}
where $w_k = \sqrt{k^2 + m^2}$. The Fourier mode of
$\varphi^{(1)}$ in Eq.~(\ref{pwescalar}) takes the form
\begin{eqnarray}
\varphi_{\textbf{k}}^{(1)} & = & 2 \bar{\varphi}_{\textbf{k}} \cos (w_k t)
. \label{solsdm1}
\end{eqnarray}
One should note that $\bar{\varphi}_\textbf{k}$ is time-independent, while ${\varphi}_\textbf{k}$ is not. 
Substituting Eq.~(\ref{solsdm1}) into Eqs.~(\ref{moesdm}) and evaluating it in Fourier space, we obtain the metric perturbations in the form of
\begin{subequations}
  \begin{eqnarray}
    h_{a   b, \textbf{k}} & = & 2 \kappa \int \frac{ {\rm d}^3 p}{(2
    \pi)^3} \left\{ \Lambda_{ab}^{c   d} p_c p_d \frac{\left(
    m^2 + p^2 - \textbf{k} \cdot \textbf{p} \right) \varphi_{\textbf{k} -
    \textbf{p}}^{(1)} \varphi_{\textbf{p}}^{(1)} + \left( \varphi_{\textbf{k} -
    \textbf{p}}^{(1)} \right)' \left( \varphi_{\textbf{p}}^{(1)} \right)'}{m^2
    k^2 + k^2 p^2 - \left( \textbf{k} \cdot \textbf{p} \right)^2} \right\} ~,
    \label{metsdm1}\\
    V_{b, \textbf{k}} & = & 4 i \kappa \int \frac{ {\rm d}^3 p}{(2 \pi)^3} \left\{
    \frac{\textbf{k} \cdot \textbf{p} \left( \textbf{k} \cdot
    \textbf{p} k_b - k^2 p_b  \right)}{k^4 \left( k^2 - 2 \textbf{k} \cdot \textbf{p}
    \right)} \left( \left( \varphi_{\textbf{k} - \textbf{p}}^{(1)} \right)'
    \varphi_{\textbf{p}}^{(1)} - \varphi_{\textbf{k} - \textbf{p}}^{(1)} \left(
    \varphi_{\textbf{p}}^{(1)} \right)' \right) \right\} ~, \\
    \psi_{\textbf{k}}^{(2)} & = & - \frac{\kappa}{2} \int \frac{ {\rm d}^3 p}{(2
    \pi)^3} \left\{ \frac{m^2 + p^2 - \textbf{k} \cdot \textbf{p}}{k^2}
    \varphi_{\textbf{k} - \textbf{p}}^{(1)} \varphi_{\textbf{p}}^{(1)} +
    \frac{1}{k^2} \left( \varphi_{\textbf{k} - \textbf{p}}^{(1)} \right)' \left(
    \varphi_{\textbf{p}}^{(1)} \right)' \right\} ~, \\
    \phi^{(2)}_{\textbf{k}} & = & - \frac{\kappa}{2} \int \frac{ {\rm d}^3 p}{(2
    \pi)^3} \left\{ \frac{m^2 k^2 - 6 \left( \textbf{k} \cdot \textbf{p}
    \right)^2 + 3 k^2 \left( p^2 + \textbf{k} \cdot \textbf{p} \right)}{k^4}
    \varphi_{\textbf{k} - \textbf{p}}^{(1)} \varphi_{\textbf{p}}^{(1)} +
    \frac{1}{k^2} \left( \varphi_{\textbf{k} - \textbf{p}}^{(1)} \right)' \left(
    \varphi_{\textbf{p}}^{(1)} \right)' \right\} ~, \nonumber\\ \label{metsdm4}
  \end{eqnarray}\label{metsdm}
\end{subequations}
where we have neglected the non-oscillatory parts. By making use of Eq.~(\ref{solsdm1}) and (\ref{metsdm}), we can expand the solution of metric perturbations at $k/m \rightarrow0$ as follows,
\begin{subequations}
  \begin{eqnarray}
    h_{a   b, \textbf{k}} & = & - \frac{2 \kappa}{m^2} \int
    \frac{ \textrm{d}^3 p}{(2 \pi)^3} \{  \Lambda^{c   d}_{a   b} p_c
    p_d \bar{\varphi}_{\textbf{k}-\textbf{p}}\bar{\varphi}_\textbf{p}\cos (w_{ \textrm{sdm}} t) \}~,\\
    V_{b, \textbf{k}} & = & \frac{4 i \kappa}{m   k^4} \int
    \frac{ \textrm{d}^3 p}{(2 \pi)^3} \left\{ \textbf{k} \cdot \textbf{p} \left( k^2
    p_b - \textbf{k} \cdot \textbf{p} k_b \right) \bar{\varphi}_{\textbf{k}-\textbf{p}}\bar{\varphi}_\textbf{p}\sin (w_{ \textrm{sdm}} t)
    \right\}~,\\
    \psi^{(2)}_{\textbf{k}} & = & \frac{\kappa}{2} \int \frac{ \textrm{d}^3 p}{(2
    \pi)^3} \left\{ \left( 1 - \frac{\left( k^2 - 2 \textbf{k} \cdot \textbf{p}
    \right)^2}{4 m^2 k^2} \right)\bar{\varphi}_{\textbf{k}-\textbf{p}}\bar{\varphi}_\textbf{p} \cos (w_{ \textrm{sdm}} t) \right\}~, \label{sdmpsi}\\
    \phi^{(2)}_{\textbf{k}} & = & \frac{\kappa}{2} \int \frac{ \textrm{d}^3 p}{(2
    \pi)^3} \Bigg\{ \left( \frac{k^4 + 12 \left( \textbf{k} \cdot \textbf{p}
    \right)^2 - 4 k^2 \left( p^2 + 2 \textbf{k} \cdot \textbf{p} \right)}{k^4} -
    \frac{\left( k^2 - 2 \textbf{k} \cdot \textbf{p} \right)^2}{4 m^2 k^2}
    \right) \nonumber \\ && \times \bar{\varphi}_{\textbf{k}-\textbf{p}}\bar{\varphi}_\textbf{p} \cos (w_{ \textrm{sdm}} t) \Bigg\}~,
  \end{eqnarray} \label{metsdm0}
\end{subequations}
where $w_\textrm{sdm}$ is given by
\begin{eqnarray}
  w_{\textrm{sdm}} & = & 2 m + \frac{k^2 + 2 p^2 - 2 \textbf{k} \cdot
  \textbf{p}}{2 m}~. \label{wsdm}
\end{eqnarray}
Compared to the leading order angular frequency $w_\textrm{sdm} \simeq 2m$ \cite{Khmelnitsky:2013lxt}, it is shifted by the order of $\mathcal{O}((k / m)^2)$. We have ignored the terms involving the low-frequency modes $w_\textrm{sdm} \simeq (k^2 - 2\textbf{k} \cdot \textbf{p}) / (2m)$ in Eqs.~(\ref{metsdm0}).  The amplitudes of the metric perturbations can be estimated as $h_{ab,\textbf{k}} \simeq \mathcal{O}((k/m)^2)$, $V_{b,\textbf{k}} \simeq \mathcal{O}(k/m)$, and $\psi^{(2)}_{\textbf{k}} \simeq \phi^{(2)}_{\textbf{k}} \simeq \mathcal{O}(1)$. It indicates that tensor and vector metric perturbations are sub-dominant.

\

\subsection{Non-adiabatic perfect fluid dark matter}

The energy-momentum tensor of the perfect fluid is given by $T_{\mu \nu} = (p + \rho) u_{\mu} u_{\nu} + g_{\mu \nu} p$, where the density $\rho$, pressure $p$, and velocity flow $u_{\mu}$ can be expanded in the form of
  \begin{eqnarray}
    \rho  =  \rho^{(1)} + \frac{1}{2} \rho^{(2)} +\mathcal{O} (3) ~, \hspace{0.4cm}
    p  =  w \rho^{(1)} + \frac{1}{2} c_{\rm s}^2 \rho^{(2)} +\mathcal{O} (3) ~,
    \hspace{0.4cm}
    u^i  =  \upsilon^{(1), i} + \frac{1}{2} \upsilon^{(2), i} +\mathcal{O} (3)
    ~.
  \end{eqnarray}
  The $w$ is the equation of state parameter, and $c_{\rm s}$ is the effective sound speed. The perfect fluid is an effective description for dark matter. For instance, the dark matter originating from primordial black holes can be described as perfect fluid \cite{Sasaki:2018dmp,Domenech:2021ztg}. Hence, the sound speed $c_\text{s}$ should be in the order of the typical velocity in our galaxy, $v_\textrm{Gal} \simeq \mathcal{O}(10^{-3})$ \cite{NANOGrav:2023hvm}.  Defining $\rho^{(0)}$ and $p^{(0)}$ is meaningless, since the background in Eq.~(\ref{met}) is Minkowski space-time

  Expanding the equations of energy-momentum tensor conservation to the first order, we obtain  
\begin{eqnarray}
  {\rho^{(1)}}'  =  0 ~,  \label{DTpf1} &&
  w{\rho^{(1)}}  =  0 ~.  \label{DTpf2}
\end{eqnarray}
We adopt the scenario where $w=0$ and $\rho^{(1)} (\neq 0)$ is stationary. The latter corresponds to the dark matter density comprising the galactic halo. It is consistent with the picture that perfect fluid dark matter can serve as a type of CDM within our galaxy.
Expanding the equations of the energy-momentum tensor conservation in the second order, we obtain
  \begin{eqnarray}
    (\upsilon_i^{(1)} \rho^{(1)})'' - c_{\rm s}^2 \partial_i \partial_j
    (\upsilon^{(1), j} \rho^{(1)})  =  0 ~, &&
    {\rho^{(2)}}' + 2 \partial_j (\upsilon^{(1), j} \rho^{(1)})  =  0 ~.
  \end{eqnarray}
Evaluating the above equation, we have
\begin{eqnarray}  
  {\rho^{(2)}}'' - c_{\rm s}^2 \Delta \rho^{(2)} & = & 0 ~.  \label{DT2pf3}
\end{eqnarray}
For non-adiabatic perfect fluid $c_{\rm s} \neq 0$, and $c_{\rm s}$ is an independent quantity from the equation of state parameter $w$. The wave equation in Eq.~(\ref{DT2pf3}) suggests the presence of acoustic oscillations of the density perturbation in our galaxy.

Evaluating Einstein field equations, we obtain the first-order equations as
\begin{eqnarray}
  2 \Delta \psi^{(1)}  =  \kappa \rho^{(1)} ~,  \label{E1} &&
  \phi^{(1)}  =  \psi^{(1)} ~, 
\end{eqnarray}
and the second-order equations in the form of
\begin{subequations}
  \begin{eqnarray}
    \frac{1}{4} (h_{ab}'' - \Delta h_{ab}) & = & \Lambda_{a
      b}^{c   d} S^{({\rm pfdm})}_{c   d} ~, 
    \label{moepf1}\\
    - \frac{1}{4} V_a' & = & \Delta^{- 1} \partial^c (\delta_a^d - \partial_a
    \Delta^{- 1} \partial^d) S^{({\rm pfdm})}_{c   d} ~, \\
    {\psi^{(2)}}'' & = & \partial^c \Delta^{- 1} \partial^d S^{({\rm pfdm})}_{c
      d} + \frac{1}{2} \kappa c_{\rm s}^2 \rho^{(2)} ~, \\
    \frac{1}{2} (\psi^{(2)} - \phi^{(2)}) & = & \frac{1}{2} \Delta^{- 1} (3
    \partial^c \Delta^{- 1} \partial^d - \delta^{c   d}) S_{c  
    d}^{({\rm pfdm})} ~,  \label{moepf4}
  \end{eqnarray} \label{moepf}
\end{subequations}
where the $S^{({\rm pfdm})}_{  b   c}$ on the right-hand side of above equation is
\begin{eqnarray}
  S^{({\rm pfdm})}_{  b   c} & = & \delta_{b   c} (4
  \psi^{(1)} \Delta \psi^{(1)} + 3 \partial_d \psi^{(1)} \partial^d
  \psi^{(1)}) - 2 \partial_b \psi^{(1)} \partial_c \psi^{(2)} - 4 \psi^{(1)}
  \partial_b \partial_c \psi^{(1)} ~.  \label{srcpf}
\end{eqnarray}
Based on Eqs.~(\ref{DTpf1}) and (\ref{E1}), we obtain a time-independent $\psi^{(1)}$, consequently leading to $S_{bc}^{({\rm pfdm})}$ in Eq.~(\ref{srcpf}) being time-independent. In this study, we focus on the oscillatory features of space-time fluctuations that could potentially be reflected in pulsar timing. Therefore, Eqs.~(\ref{moepf}) are evaluated by neglecting the non-oscillatory parts, namely,
  \begin{eqnarray}
    h_{ab}'' - \Delta h_{ab}  =  0 ~, &&  V_a'  =  0 ~, \hspace{0.5cm}
    {\psi^{(2)}}''  =  \frac{1}{2} \kappa c_{\rm s}^2 \rho^{(2)} ~,  \hspace{0.5cm}
    \psi^{(2)}  =  \phi^{(2)} ~. \label{moe:pf}
  \end{eqnarray}
  It shows that only the Newtonian potential perturbation and the curvature perturbation can be generated by the oscillating perfect fluid dark matter.
  Based on Eqs.~(\ref{DT2pf3}), (\ref{moepf}) and (\ref{moe:pf}), $\psi^{(2)}$ should satisfy the wave equation in the form of Eq.~(\ref{DT2pf3}). Therefore, by solving the equations, one can obtain Fourier modes of $\psi^{(2)}$ and $\phi^{(2)}$ as
  \begin{eqnarray}
    \phi^{(2)}_{\textbf{k}} = \psi^{(2)}_{\textbf{k}} & = & 2
    \bar{\psi}_{\textbf{k}} \cos (c_{\rm s} k   t) ~.  \label{metpfdm}
  \end{eqnarray}
  Above metric perturbations also take the form of a plane wave expansion similar to Eq.~(\ref{solsdm1}). 
  The $\bar{\psi}_\textbf{k}$ is time-independent, while $\phi_\textbf{k}$ and $\psi_\textbf{k}$ are not. 
  
  The scalar dark matter density should be derived from the time average of $\rho^{(2)}$ due to $\rho^{(1)}=0$. Consequently, the amplitude of the metric perturbations can be expressed in terms of the dark matter density \cite{Khmelnitsky:2013lxt}. However, no such relation exists for non-adiabatic perfect fluid dark matter. In this case, the amplitude in Eq.~(\ref{metpfdm}) and the corresponding dark matter density are independent quantities.

\section{Timing residuals and angular correlations in pulsar timing arrays\label{III}}

The radio beam from a pulsar propagates along geodesics in space. In the presence of space-time fluctuations between the pulsar and the earth, it can leave an imprint along the propagation of the radio beam, consequently affecting the pulsar timing. It can be formally derived from the perturbed geodesic equations, namely, 
\begin{eqnarray}
  \delta P^{\mu} \partial_{\mu} \bar{P}^{\nu} + \bar{P}^{\mu} \partial_{\mu}
  \delta P^{\nu} + \eta^{\nu \rho} \left( \partial_{\mu} \delta g_{\lambda
  \rho} - \frac{1}{2} \partial_{\rho} \delta g_{\mu \lambda} \right)
  \bar{P}^{\mu} \bar{P}^{\lambda} & = & 0 ~, \label{pergeo}
\end{eqnarray}
where the 4-velocity is expanded as $P^{\mu} = \bar{P}^{\mu} + P^{(1), \mu} +({1}/{2}) P^{(2), \mu}$. As shown in previous sections, the time-dependent metric perturbations are all second order. Thus, we have $P^{(1), \mu} = 0$, $\delta P^{\mu} = P^{(2), \mu}$, and $\delta g_{\lambda \sigma} = \delta g_{\mu \nu}^{(2)}$. Associating Eq.~(\ref{met}) with Eq.~(\ref{pergeo}), we obtain the relative variation of the temporal component of the 4-velocity, namely,
\begin{eqnarray}
  \frac{\delta P^0}{\bar{P}^0}  &=& \psi^{(2)} - \phi^{(2)} + \frac{1}{2} \hat{n}^i \hat{n}^j h_{i   j} \nonumber \\ && +
\int {\rm d} t \left\{ \hat{n}^i \partial_j \left( \phi^{(2)} \left( t,
\textbf{x} (t) \right) + \psi^{(2)} \left( t, \textbf{x} (t) \right) +
\hat{n}^i V_i \left( t, \textbf{x} (t) \right) - \frac{1}{2} \hat{n}^i
\hat{n}^j h_{i   j} \left( t, \textbf{x} (t) \right) \right) \right\}~, \label{dp}
\end{eqnarray}
where we have introduced the direction vector $\hat{n}^i \equiv - {\rm d} x^i / {\rm d} t = -
\bar{P}^i / \bar{P}^0$, representing the location of a pulsar on the sky.
The response of PTA observations to space-time fluctuations can be quantified by
\begin{eqnarray}
  z & = & \frac{1}{2} \left. \frac{\delta P^0}{\bar{P}^0} \right|^O_E \nonumber \\
  & = & \left. \frac{1}{2} \left( \psi^{(2)} - \phi^{(2)} - \frac{1}{2}
  \hat{n}^i \hat{n}^j h_{i  j} \right) \right|^O_E \nonumber\\
  &  & + \frac{1}{2} \int^{t_O}_{t_E} {\rm d} \tau \left\{ \hat{n}^l
  \partial_l \left( \psi^{(2)} \left( \tau, \textbf{x} (\tau) \right) +
  \phi^{(2)} \left( \tau, \textbf{x} (\tau) \right) + \hat{n}^i V_i \left(
  \tau, \textbf{x} (\tau) \right) - \frac{1}{2} \hat{n}^i \hat{n}^j h_{i
   j} \left( \tau, \textbf{x} (\tau) \right) \right) \right\}~, \label{zgen}
\end{eqnarray}
where the emission event occurs at $x^{\mu}_O = (t, 0)$ and the reception event occurs at $x^{\mu}_E = (t - L, L \hat{n}^i)$. The trajectory of the radio beam is given by $\textbf{x}(\tau) = \hat{n} (t - \tau)$, with $L$ representing the distance from the earth to the pulsar. The quantity $z$ describes the time shift of radio pulses from a pulsar, because $z\simeq\delta\nu/\nu=-\delta T/T$, where $T$ is rotation period of the pulsar \cite{Maggiore:2018sht}.

In this study, we consider two distinct types of dark matter sources in PTA observations. The first type is deterministic source, where the dark matter oscillates with a specific frequency and propagation direction. Both amplitude and phase are set to be deterministic quantities. The second type is  stochastic source, where the strain or amplitude is assumed as a Gaussian stochastic variable. Therefore, the two-point correlation of the strain is used to statistically characterize the properties of the dark matter. In subsequent sections, we will study these two types of sources in the contexts of scalar dark matter and perfect fluid dark matter in PTAs.

\subsection{Ultralight scalar dark matter \label{IIIA}}
Based on the quantity $z$ in Eq.~(\ref{zgen}) and metric perturbation given in Eqs.~(\ref{metsdm0}),
one might find that the subleading-order correction of $\mathcal{O}(k / m)$ in $z$ should come from the Newtonian potential $\phi$ and curvature perturbations $\psi$. This is because $\partial_j\phi$ and $\partial_j\psi$ are of order $\mathcal{O}(k/m)$, while $h_{ab}$ and $\partial_j V_b$ are of order $\mathcal{O}((k/m)^2)$. Thus, we can restrict our attention to the scalar metric perturbation $\phi$ and $\psi$.
For illustration, we rewrite these metric perturbations as 
\begin{eqnarray}
  \Psi \left( t, \textbf{x} \right) & = & \int \frac{ {\rm d}^3 k}{(2 \pi)^3}
  \left\{ e^{i \textbf{k} \cdot \textbf{x}} \int \frac{ {\rm d}^3 p}{(2 \pi)^3}
  \left\{ 2 \cos (w_\text{sdm}   t) I_{\Psi} \left( \left| \textbf{k} -
  \textbf{p} \right|, p, k \right) \bar{\varphi}_{\textbf{k} - \textbf{p}}
  \bar{\varphi}_{\textbf{p}} \right\} \right\}~, \label{Psi}
\end{eqnarray}
where $\Psi = \psi^{(2)}$ or $\phi^{(2)}$, $w_\text{sdm}$ has been given in Eq.~(\ref{wsdm}). Based on Eqs.~(\ref{metsdm0}), the function $I_{\Psi} \left( \left| \textbf{k} -\textbf{p} \right|, p, k \right)$ can be given by 
\begin{eqnarray}
  I_{\psi} \left( \left| \textbf{k} - \textbf{p} \right|, p, k \right) = 
  \frac{\kappa}{2}~, & &
  I_{\phi} \left( \left| \textbf{k} - \textbf{p} \right|, p, k \right)  = 
  \frac{\kappa}{2} \left( 1 + \frac{12 \left( \textbf{k} \cdot \textbf{p}
  \right)^2 - 4 k^2 \left( p^2 + 2 \textbf{k} \cdot \textbf{p} \right)}{k^4}
  \right)~.
\end{eqnarray}
Following the approach in Ref.~\cite{Maggiore:2007ulw}, we transform the metric perturbations in Eq.~(\ref{Psi}) into expansions in terms of frequency $f$ and propagation direction of the wave-like dark matter, $\hat{k}$, namely,
\begin{eqnarray}
  \Psi & = & \int_{- \infty}^{\infty}  {\rm d} f   \int  {\rm d} \Omega
  \left\{ \int \frac{ {\rm d} k  {\rm d}^3 p}{(2 \pi)^5} \left\{ k^2 \delta (2 \pi
  | f | - w_\text{sdm}) I_{\Psi} \left( \left| \textbf{k} - \textbf{p} \right|, p, k
  \right) \bar{\varphi}_{\textbf{k} - \textbf{p}} \bar{\varphi}_{\textbf{p}}
  e^{- i \left( 2 \pi f   t - \textbf{k} \cdot \textbf{x} \right)}
  \right\} \right\}~. \label{Phi2}
\end{eqnarray}
In Appendix~\ref{App1}, we provide the calculation details with a simple example.
In the subsequent parts, we will study observable $z$ in PTAs from deterministic or
stochastic scalar dark matter based on Eq.~(\ref{Phi2}).

\subsubsection{Time residuals from determinstic sources}

Considering deterministic scalar dark matter propagating in a specific direction and at a specific frequency, we can set $\bar{\varphi}_\textbf{k} = (2\pi)^3\mathcal{A}_\varphi \delta^3 ( \textbf{k} - \textbf{k}_{*} )$, where $\textbf{k}_{*}$ is the characteristic wavenumber \cite{Maggiore:2007ulw}. In this case, the metric perturbation in Eq.~(\ref{Phi2}) reduces to
\begin{eqnarray}
  \Psi & = & \mathcal{A}_{\varphi}^2 I_{\Psi} (k_{\ast}, k_{\ast}, 2 k_{\ast})
  \cos ( 2 \pi   f_\ast  t - 2   \textbf{k}_{\ast} \cdot
  \textbf{x}  )~,
\end{eqnarray}
where we only consider the real and positive frequency part of $\Psi$. The frequency is determined by $f_\ast = (2 \pi)^{-1} \left(2 m + {k_{\ast}^2}/{m}\right)$ with the dispersion relation, $w_\text{sdm} = 2 \pi |f|$. The quantity $I_{\Psi}(k_{\ast}, k_{\ast}, 2 k_{\ast})$ in Eq.~(\ref{Phi2}) can be given  explicitly, namely, $I_{\psi}(k_{\ast}, k_{\ast}, 2 k_{\ast}) = -I_{\phi}(k_{\ast}, k_{\ast}, 2 k_{\ast}) = {\kappa}/{2}$. This implies that $\psi^{(2)} = -\phi^{(2)}$ for the deterministic scalar dark matter. With above results, the $z$ can be expressed as
\begin{eqnarray}
  z & = & \psi^{(2)} (t, 0) - \psi^{(2)} (t - L, L \hat{n}^i) \nonumber \\
  & = & \frac{\kappa \mathcal{A}_{\varphi}^2}{2} (\cos (2 \pi   f_\ast
    t) - \cos (2 \pi   f_\ast(t - L) - 2 k_{\ast} L (\hat{k}_{\ast} \cdot
  \hat{n})))~.
\end{eqnarray} 
It shows that the time integration term in Eq.~(\ref{zgen}) vanishes due to $\psi^{(2)} = -\phi^{(2)}$. In the limit as $k_\ast/(2\pi f_\ast) \rightarrow 0$, this result reduces to that obtained in Ref.~\cite{Khmelnitsky:2013lxt}\footnote{One can let $\mathcal{A}_\varphi \to ({1}/{2})\mathcal{A}_{\varphi^{(1)}}$, where $\mathcal{A}_{\varphi^{(1)}}$ is the amplitude of $\varphi^{(1)}_\textbf{k}$, in a consistent convention used in Ref.~\cite{Khmelnitsky:2013lxt}. Our result for $z$ seems to be double that of Eq.~(3.3) in Ref.~\cite{Khmelnitsky:2013lxt}, due to the setup of the perturbed metric in Eq.~(\ref{met}) and rigorously considering $\phi^{(2)}$ left in $z$.
}.

Following the approaches used in Ref.~\cite{Khmelnitsky:2013lxt}, we further calculate the time residual $R$ by integrating over time $t$ and neglecting the non-oscillatory part, namely, 
\begin{eqnarray}
  R^\text{(sdm)} & = & \mathcal{R}^\text{(sdm)} \cos (2 \pi f_\ast  t - L (\pi f_\ast(1 + 2 \alpha
  \hat{k}_\ast \cdot \hat{n})))~,
\end{eqnarray}
where $\alpha = k_{\ast} / (2 \pi f_\ast)$ and timing residual amplitude is 
\begin{eqnarray}
  \mathcal{R}^\text{(sdm)} & = & \frac{\kappa \mathcal{A}_{\varphi}^2}{2 \pi f_\ast} \sin (\pi f_\ast  ( 2 \alpha \hat{k}_\ast \cdot \hat{n}-1))~.
\end{eqnarray}
It shows that $\mathcal{R}^\text{(sdm)}$ is direction-dependent, due to the term $2 \alpha \hat{k}_\ast \cdot \hat{n}$ attributed to the subleading-order correction of $\mathcal{O}(k_\ast/m)$.

\begin{figure}
  \includegraphics[width=0.85\linewidth]{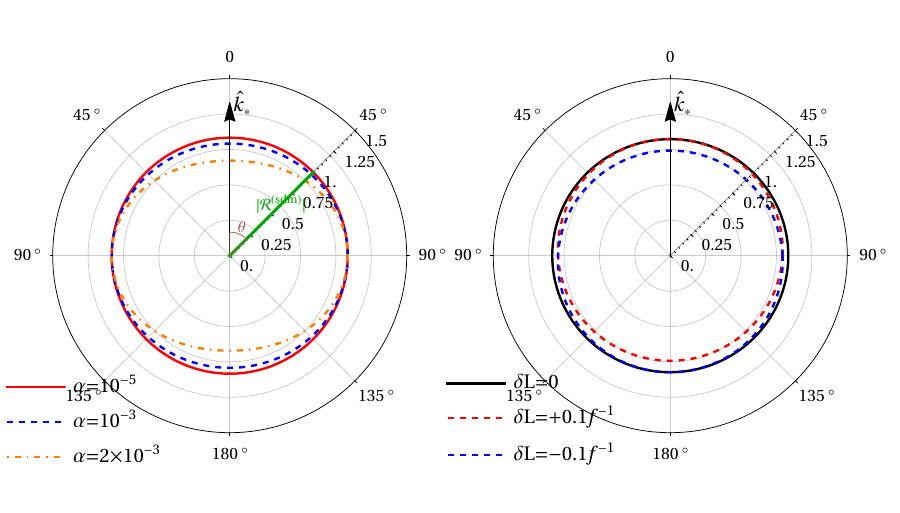}
  \caption{The curves of $| \mathcal{R}^\text{(sdm)}|$ as a function of $\theta \equiv \arccos (\hat{k}_\ast \cdot \hat{n})$ in polar coordinates. For illustration, we let  ${\kappa \mathcal{A}_{\varphi}^2}/{(2 \pi f_\ast)}=1$ for the curves. Left panel: the curves of $| \mathcal{R}^\text{(sdm)}|$ is plotted for selected $\alpha$ with fixed $fL=50.5$, and right panel: the curves of $| \mathcal{R}^\text{(sdm)}|$ for selected $L=(50.5+\delta L)f^{-1}$ with fixed $\alpha=0.0005$. \label{F1}} 
\end{figure}

Figure~\ref{F1} shows the directional dependence of $|\mathcal{R}^\text{(sdm)}|$ in polar coordinates. In the case of $\alpha \rightarrow 0$, shown in the left panel of Figure~\ref{F1}, the curves reduce to a circle, indicating no directional dependence. For $fL \sim \mathcal{O}(50)$ in PTA observations \cite{Maggiore:2018sht}, the directional dependence becomes distinguishable even for small values of $\alpha$.  Besides, variations in the distance of a pulsar, $\delta L$, can also affect the directional dependence of the timing residuals. It suggests that the subleading-order correction of $k/m$ is non-negligible in practice, since the $2\alpha (\simeq k/m \approx 10^{-3})$ is of the order of the typical velocity in the galaxy \cite{Khmelnitsky:2013lxt}. We will further study the impact of $L$ on the timing residual amplitude in Section~\ref{IV}.

\subsubsection{Angular correlations from stochastic sources}

For stochastic scalar dark matter, the $z$ in Eq.~(\ref{zgen}) can be evaluated by making use of Eq.~(\ref{Phi2}), namely,
\begin{eqnarray}
  z & = & \frac{1}{2} \int_{- \infty}^{\infty} {\rm d} f   \int {\rm d}
  \Omega \Bigg\{ \int \frac{{\rm d} k {\rm d}^3 p}{(2 \pi)^5} \Big\{ k^2 \delta
  (2 \pi | f | - w_\text{sdm}) \bar{\varphi}_{\textbf{k} - \textbf{p}}
  \bar{\varphi}_{\textbf{p}}  \nonumber \\ 
  & & \times \mathcal{I} \left( f, \hat{n} ; \textbf{k},
  \left| \textbf{k} - \textbf{p} \right|, p \right) e^{- 2 \pi i   f
    t} \left( 1 - e^{i   L \left( 2 \pi f + \textbf{k} \cdot
  \hat{n} \right)} \right) \Big\} \Bigg\}~, \label{zsdm2}
\end{eqnarray}
where
\begin{eqnarray}
  \mathcal{I} \left( f, \hat{n} ; \textbf{k}, \left| \textbf{k} - \textbf{p}
  \right|, p \right) & \equiv & I_{\psi} \left( \left| \textbf{k} - \textbf{p}
  \right|, p, k \right) \left( 1 - \frac{\textbf{k} \cdot \hat{n}}{2 \pi f +
  \textbf{k} \cdot \hat{n}} \right) \nonumber \\ && - I_{\phi} \left( \left| \textbf{k} -
  \textbf{p} \right|, p, k \right) \left( 1 + \frac{\textbf{k} \cdot
  \hat{n}}{2 \pi f + \textbf{k} \cdot \hat{n}} \right)~.
\end{eqnarray}
The $\mathcal{I} \left( f, \hat{n} ; \textbf{k}, \left| \textbf{k} - \textbf{p}
\right|, p \right)$ is shown to be direction-dependent, because of $I_\psi\neq-I_\phi$.
For stochastic source, the Fourier mode $\bar{\varphi}_\textbf{k}$ in Eq.~(\ref{zsdm2}) is considered as a stochastic variable. Its statistical properties can be described by the spatial correlation of $\varphi_\textbf{k}$ in the form of
\begin{eqnarray}
  \left\langle \bar{\varphi}_{\textbf{k}'} \bar{\varphi}_{\textbf{k}}
  \right\rangle & = & (2 \pi)^3 \delta \left( \textbf{k} + \textbf{k}' \right)
  P_{\varphi} (k)~. \label{cor:varphi}
\end{eqnarray}
where $P_\varphi(k)$ is the power spectrum of the scalar field, and it is assumed to be Gaussian, stationary and isotropic. One can also introduce a dimensionless power spectrum $\mathcal{P}_\varphi(k)\equiv (2\pi^2/k^3)P_\varphi(k)$.
By making use of Eq~(\ref{zsdm2}) and (\ref{cor:varphi}), the correlation of $z$ for a pair of pulsars can be obtained, namely,
\begin{eqnarray}
  \langle z_A z_B^{\ast} \rangle & = & \int_{- \infty}^{\infty} {\rm d} f    \int {\rm d} \Omega \int \frac{{\rm d} k {\rm d}^3 p}{(2 \pi)^5}  \Bigg\{ k^2 \delta (2 \pi | f | - w_\text{sdm}) P_{\varphi} \left( \left| \textbf{k}  - \textbf{p} \right| \right) P_{\varphi} (p) \nonumber \\
  &  & \mathcal{I} \left( f, \hat{n}_\text{A} ; \textbf{k}, \left| \textbf{k} -  \textbf{p} \right|, p \right) \left( \mathcal{I} \left( f, \hat{n}_\text{B} ;  \textbf{k}, \left| \textbf{k} - \textbf{p} \right|, p \right) +\mathcal{I}  \left( f, \hat{n}_\text{B} ; \textbf{k}, p, \left| \textbf{k} - \textbf{p} \right|  \right) \right) \nonumber \\ && \left( 1 - e^{i   L_\text{A} \left( 2 \pi f + \textbf{k} \cdot
  \hat{n}_\text{A} \right)} \right) \left( 1 - e^{- i   L_\text{B} \left( 2 \pi f +
  \textbf{k} \cdot \hat{n}_\text{B} \right)} \right) \Bigg\}~. \label{cor:zsdmAB}
\end{eqnarray}
Similar to the studied on stochastic GW background, one can rewrite the above correlation in the form of
$\langle z_A z_B^{\ast} \rangle = \frac{1}{2} \int_{- \infty}^{\infty}
{\rm d} f   \{ S (f) \Gamma^\text{(sdm)} (f, \theta_\text{AB}) \}$ \cite{Romano:2016dpx,Maggiore:2018sht}, where $S
(f)$ is the spectral density of signals, and $\Gamma^\text{(sdm)} (f,
\theta_\text{AB})$ is overlap reduction function (ORF). Based on Eq.~(\ref{cor:zsdmAB}), we can obtain ORFs in the form of
\begin{eqnarray}
  \Gamma^{(\textrm{sdm})} (f, \theta_{\textrm{AB}}) & = & \frac{2}{S (f)} \int
  \frac{\textrm{d} \Omega}{4 \pi} \int_{k_{\textrm{Gal}}}^{\infty} \frac{\textrm{d}
  k}{k} \int_0^{\infty} \textrm{d} \upsilon \int_{| 1 - \upsilon |}^{1 + \upsilon}
  \textrm{d} u \Bigg\{ \nonumber \\ && \frac{1}{2   u^2 \upsilon^2} \mathcal{P} (k  
  u) \mathcal{P} (k \upsilon) \delta \left( | f | - \frac{1}{2 \pi} \left( 2 m
  + \frac{k^2 (u^2 + \upsilon^2)}{2 m} \right) \right) \nonumber \\
  &  & \times \mathcal{I} \left( f, \hat{n}_{\textrm{A}} ; \textbf{k}, k  
  u  , k   \upsilon \right) \left( \mathcal{I} \left( f,
  \hat{n}_{\textrm{B}} ; \textbf{k}, k   u  , k   \upsilon
  \right) +\mathcal{I} \left( f, \hat{n}_{\textrm{B}} ; \textbf{k}, k   u
   , k   \upsilon \right) \right) \nonumber \\ &&  \times (1 - e^{i   L (2 \pi f
  + k   \hat{k} \cdot \hat{n}_A)}) (1 - e^{- i   L (2 \pi f + k
  \hat{k} \cdot \hat{n}_B)}) \Bigg\}~, \label{ORFsdm}
\end{eqnarray}
where we have used variable substitution, $u = \left| \textbf{k} - \textbf{p} \right| /k$, $\upsilon = p / k$ and
The $S (f)$ is set to be spectral density of curvature perturbations, namely,
\begin{eqnarray}
  S (f) & = & \kappa^2 \int \textrm{d} \Omega \int \frac{\textrm{d} k \textrm{d}^3 p}{(2
  \pi)^5} \left\{ k^2 \delta (2 \pi | f | - w_s) P_{\varphi} \left( \left|
  \textbf{k} - \textbf{p} \right| \right) P_{\varphi} (p) \right\} \nonumber \\
  & = & \kappa^2 \int \textrm{d} \Omega \int_{k_{\textrm{Gal}}}^{\infty}
  \frac{\textrm{d} k}{k} \int_0^{\infty} \textrm{d} \upsilon \int_{| 1 - \upsilon
  |}^{1 + \upsilon} \textrm{d} u \Bigg\{ \nonumber \\ &&  \frac{1}{8 \pi k   u^2 \upsilon^2}
  \mathcal{P} (k   u) \mathcal{P} (k \upsilon) \delta \left( | f | -
  \frac{1}{2 \pi} \left( 2 m + \frac{k^2 (u^2 + \upsilon^2)}{2 m} \right) 
  \right) \Bigg\}~. \label{Sfsdm}
\end{eqnarray}
There is a large-scale cutoff $k_\text{Gal}$ in Eqs.~(\ref{ORFsdm}) and (\ref{Sfsdm}), because we are only interested in the oscillating scalar dark matter within our galaxy. 
It is found that the ORFs in Eq.~(\ref{ORFsdm}) depend on the power spectrum of scalar dark matter, $\mathcal{P}_{\varphi}(k)$. This differs from results concerning stochastic gravitational waves, where the corresponding ORFs are independent of the power spectrum of GWs \cite{Romano:2016dpx,Maggiore:2018sht}.

As reported by the NANOGrav collaboration, there are indications of an additional signal confined to a narrow frequency range near 4 nHz, besides the stochastic GW \cite{NANOGrav:2023gor}. Therefore, we consider the monochromatic power spectrum of scalar dark matter, given by $\mathcal{P}_\varphi (k) = A k_{\ast} \delta (k - k_{\ast})$. 
Since it is expected that $f\simeq2m/(2\pi)$ is within PTA frequency band, and the typical velocity in the galaxy is $v \simeq k_\ast/m\simeq0.001$, it suggests that $k_\ast \simeq0.3\text{kpc}^{-1}\simeq (26.8/19)k_\text{Gal}$.
In this case, the ORFs in Eq.~(\ref{ORFsdm}) reduce to the form of
\begin{eqnarray}
  \Gamma^{(\textrm{sdm})} (f, \theta_{\textrm{AB}}) & = & \frac{2}{(4-\epsilon^2)\kappa^2} \int
  \frac{\textrm{d} \Omega}{4 \pi} \int_{\epsilon}^2 \textrm{d} \tilde{k} \Big\{ \tilde{k}
  \mathcal{I} (f, \hat{n}_{\textrm{A}} ; k_{\ast} \hat{k}, k_{\ast}  ,
  k_{\ast}) \mathcal{I} (f, \hat{n}_{\textrm{B}} ; k_{\ast} \hat{k}, k_{\ast}
   , k_{\ast}) \nonumber \\ && \times (1 - e^{i   2 \pi f   L_{\textrm{A}} (1 +
  \alpha \tilde{k}   \hat{k} \cdot \hat{n}_A)}) (1 - e^{- i   2
  \pi f   L_{\textrm{B}} (+ \alpha \tilde{k} \hat{k} \cdot \hat{n}_B)}) \Big\}~,
\end{eqnarray}
where we have $\tilde{k}\equiv k/k_\ast$, $\epsilon\equiv k_\text{Gal}/k_\ast\simeq0.7$, $\alpha \equiv k_{\ast} / (2 \pi f)$ and   
\begin{eqnarray}  
  S (f) & = & \frac{4-\epsilon^2}{2} \kappa^2 A^2 \delta \left( | f | - \frac{1}{2 \pi}
  \left( 2 m + \frac{k_{\ast}^2}{m} \right) \right)~,\\
  \mathcal{I} (f, \hat{n} ; k_{\ast} \hat{k}, k_{\ast}  , k_{\ast}) & =
  & \frac{\kappa}{2} \left( 1 + \frac{4}{\tilde{k}^2} + \alpha \left(
  \frac{4}{\tilde{k}} - \tilde{k} \right) \left( \frac{\hat{k} \cdot
  \hat{n}}{1 + \alpha \tilde{k} \hat{k} \cdot \hat{n}} \right) \right)~.
\end{eqnarray} 
As expected, the spectral density $S(f)$ is monochromatic with the ansatz of $\mathcal{P}_\varphi(k)$. For $\alpha = 0$, the $\Gamma^\text{(sdm)}(f,\theta_\text{AB})$ is shown to be independent of $\theta_\text{AB}$. 

\begin{figure}
  \includegraphics[width=0.9\linewidth]{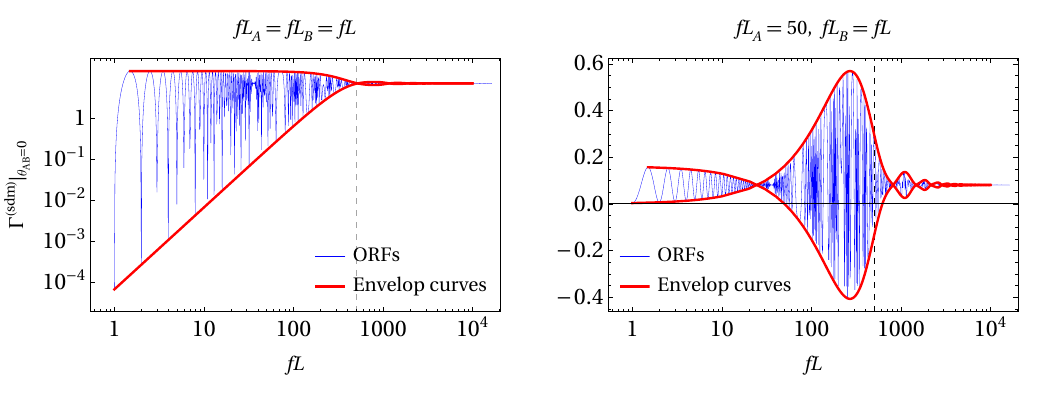}
  \caption{
    Normalized ORFs originating from scalar dark matter at $\theta_\text{AB}=0$ as a function of $fL$ for a fixed $\alpha = 0.001$. In the left panel, we set $fL_\text{A} = fL_\text{B} = fL$, while in the right panel, $fL_\text{A}$ is fixed and $fL_\text{B} = fL$. The envelope curves of the ORFs are shown by the red curves. The gray dashed lines represent $fL = 1/(2\alpha)$.
    \label{F2}}
\end{figure}
\begin{figure}
  \includegraphics[width=0.9\linewidth]{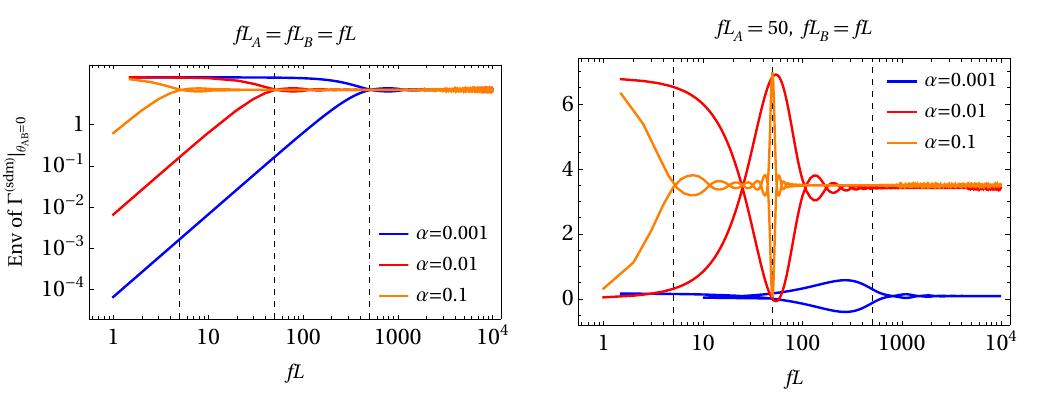}
  \caption{Envelope curves of the ORFs at $\theta_\text{AB}=0$ as a function of $fL$ for selected values of $\alpha$. In the left panel, we set $fL_\text{A} = fL_\text{B} = fL$, while in the right panel, $fL_\text{A}$ is fixed and $fL_\text{B} = fL$. The gray dashed lines represent $fL = 1/(2\alpha)$.
    \label{F3}}
\end{figure}
\begin{figure}
  \includegraphics[width=0.6\linewidth]{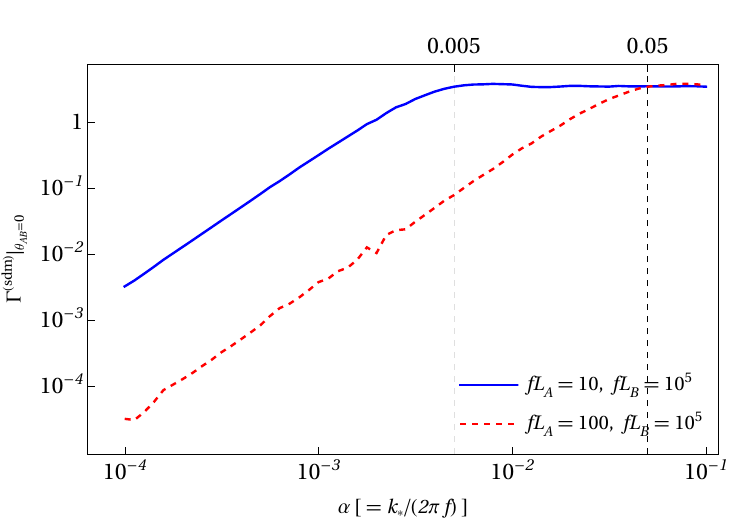}
  \caption{ Normalized ORFs at $\theta_\text{AB}=0$ as function of $\alpha$ for given $fL_\text{A}$ and a large $fL_\text{B}$.
    \label{F4}}
\end{figure}
\begin{figure}
  \includegraphics[width=1\linewidth]{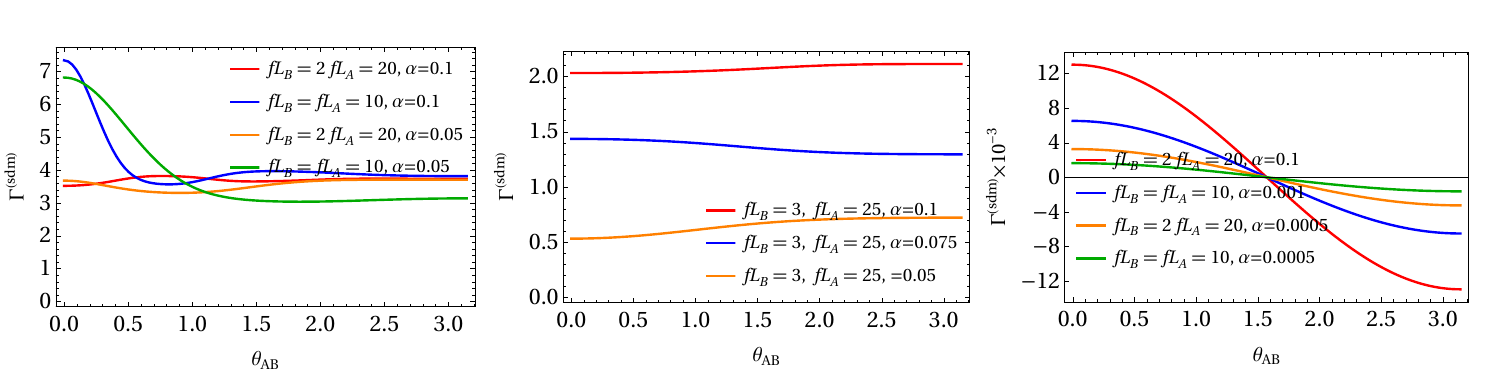}  
  \caption{The angular correlations of the $z$ quantified by the ORFs for given $fL_\text{A}$, $fL_\text{B}$, and $\alpha$. Specifically, we set left panel: $2\alpha fL_\text{A}, 2\alpha fL_\text{B}\gtrsim1$, medium panel: $2\alpha fL_\text{A}<1$ and  $2\alpha fL_\text{B}\gtrsim1$, and right panel: $2\alpha fL_\text{A}, 2\alpha fL_\text{B}<1$. \label{F5}}
\end{figure}
Figure~\ref{F2} shows the ORFs at $\theta_\text{AB}=0$ varying with the distance of pulsars. The values of $\Gamma^\text{(sdm)}|_{\theta_\text{AB}=0}$ oscillate with $fL$ when $fL \lesssim \mathcal{O}(10^3)$. In this regime, small variations in $L$ can significantly affect the ORFs. To determine the conditions under which the two envelope curves converge to a single one in Figure~\ref{F2}, we present the envelope curves of ORFs for selected $\alpha$ values in Figure~\ref{F3}. It shows that $\Gamma^\text{(sdm)}|_{\theta_\text{AB}=0}$ becomes insensitive to $fL$ when either $fL_\text{A}$ or $fL_\text{B}$ exceeds $1/(2\alpha)$. As shown in the right panel of Figure~\ref{F3}, when the values of $fL_\text{A}$ and $fL_\text{B}$ approach each other, $\Gamma^\text{(sdm)}|_{\theta_\text{AB}=0}$ can either double or vanish. This feature is also observed in studies of stochastic gravitational waves \cite{Hu:2022ujx}. 
Figure~\ref{F4} shows $\Gamma^\text{(sdm)}|_{\theta_\text{AB}=0}$ as a function of $\alpha$ for a finite $fL_\text{A}$ and a large $fL_\text{B}$. It is found that $\Gamma^\text{(sdm)}|_{\theta_\text{AB}=0}$ increases with $\alpha$ until both $fL_\text{A}$ and $fL_\text{B}$ exceed $1/(2\alpha)$. 
Therefore, from these facts, we can identify three typical ORFs. From left panel to right panel in Figure~\ref{F5}, the first is the ORF insensitive to small variations in $L$, the second is sensitive to $\delta L_\text{A}$ with $L_\text{A} < L_\text{B}$, and the third is sensitive to both $\delta L_\text{A}$ and $\delta L_\text{B}$. Given the PTA frequency band and the current knowledge of dark matter ($\alpha \simeq \mathcal{O}(10^{-3})$), the scalar dark matter in our galaxy should to be described by the third type of ORFs. It might challenge the observability of scalar dark matter with PTAs, as any intrinsic variation in the distance, $\delta L$, could affect PTA outputs.

\

\subsection{Perfect fluid dark matter}
 
As shown in Section~\ref{II}, only the Newtonian potential and curvature perturbations can be generated by the oscillating perfect fluid dark matter. Therefore, we restrict our attention to the Newtonian potential and curvature perturbations in Eq.~(\ref{zgen}). Following the scenario used in Eq.~(\ref{Phi2}), we also express the metric perturbations in terms of $(f,\hat{k})$, namely,
\begin{eqnarray}
  \psi^{(2)} & = & \int \frac{ {\rm d}^3 k}{(2 \pi)^3} \left\{ 2 \bar{\psi}_\textbf{k} \cos (c_\text{s} k
    t) e^{i \textbf{k} \cdot \textbf{x}} \right\} \nonumber\\
  & = & \int_{- \infty}^{\infty}  {\rm d} f \int  {\rm d} \Omega \left\{ e^{- 2
  \pi   i   f   \left( t - \frac{| f |}{c_\text{s} f} \hat{k}
  \cdot \textbf{x} \right)} \left( \frac{f}{c_\text{s}} \right)^2 \left.
  \bar{\psi}_{\textbf{k}} \right|_{k = \frac{2 \pi | f |}{c_\text{s}} } \right\}~, \label{phipf}
\end{eqnarray}
and we have $\phi^{(2)} = \psi^{(2)}$ shown in Eqs.~(\ref{moe:pf}).
One can also introduce $\tilde{\psi} (f, \hat{k}) \equiv (f /
c_\text{s})^2 \bar{\psi}_{\textbf{k}}|_{k = 2 \pi | f | / c_\text{s}}$ as presented in Appendix~\ref{App1}. By making use of Eq.~(\ref{zgen}) and (\ref{phipf}), we evaluate the  $z$ in Eq.~(\ref{zgen}) to be
\begin{eqnarray}
  z & = & \int^{t_O}_{t_E}  {\rm d} \tau \left\{ \hat{n}^l \partial_l \psi^{(2)}
  \left( \tau, \textbf{x} (\tau) \right) \right\}\nonumber\\
  & = & \int_{- \infty}^{\infty}  {\rm d} f \int  {\rm d} \Omega \left\{ \left(
  \frac{f}{c_\text{s}} \right)^2 \left. \bar{\psi}_{\textbf{k}} \right|_{k = \frac{2 \pi |
  f |}{c_\text{s}}} \left( - \frac{\hat{k} \cdot \hat{n}}{\frac{c_\text{s} f}{| f |} +
  \hat{k} \cdot \hat{n}} \right) e^{- 2 \pi i   f   t} \left( 1
  - e^{2 \pi i   f   L \left( 1 + \frac{| f |}{c_\text{s} f} \hat{k}
  \cdot \hat{n} \right)} \right) \right\}~. \label{zpfdm}
\end{eqnarray}
Because of $\psi^{(2)} = \phi^{(2)}$, there are only integration parts in Eq.~(\ref{zgen}) left. This is differed from the results for scalar dark matter. Here, we quantify the fluctuations of perfect fluid dark matter using curvature perturbation $\bar{\psi}_{\textbf{k}}$, as shown in Eq.~(\ref{zpfdm}). An alternative approach might be utilization of density contrast to describe the oscillating dark matter. However, both approaches yield consistent results in this context, because $\psi^{(2)}{}'' \propto \rho^{(2)}$, as shown in Eqs.~(\ref{moe:pf}). In the subsequent parts, we will investigate the $z$ originating from perfect fluid dark matter using the approaches presented in Section~\ref{IIIA}.

\subsubsection{Timing residuals from deterministic sources}
Suppose that there is deterministic oscillating perfect fluid dark matter propagating with a typical wavenumber $\textbf{k}_\ast$. Namely, one can set $\bar{\psi}_{\textbf{k}} \simeq (2 \pi)^3 \mathcal{A}_{\psi} \delta \left( \textbf{k} - \textbf{k}_{\ast} \right)$. In this case, we evaluate the real and positive frequency part of $z$ in Eq.~(\ref{zpfdm}) as
\begin{eqnarray}
  z & = & - \frac{\mathcal{A}_{\psi} c_\text{s} \hat{k}_{\ast} \cdot \hat{n}}{c_\text{s} +
  \hat{k}_{\ast} \cdot \hat{n}} (\cos (2 \pi   f_{\ast}   t) - \cos (2
  \pi f_{\ast}   (t - L) - 2 \pi f   L   c_\text{s}^{- 1}
  \hat{k}_{\ast} \cdot \hat{n}))~, \label{zpfdm2}
\end{eqnarray}
where $f_{\ast}\equiv c_\text{s}k_\ast/(2\pi)$. Using Eq.~(\ref{zpfdm2}), we calculate timing residuals $R^\text{(pfdm)}$ via integrating over time $t$ and neglecting the non-oscillatory part, namely,
\begin{eqnarray}
  R^{\text{(pfdm)}} & = & \mathcal{R}^{\text{(pfdm)}} \cos (2 \pi f_{\ast}   t - \pi f_{\ast}   L (1 + c_\text{s}^{-
  1} \hat{k}_{\ast} \cdot \hat{n}))~,
\end{eqnarray}
where timing residual amplitude is
\begin{eqnarray}
  \mathcal{R}^{\text{(pfdm)}} & = & - \frac{\mathcal{A}_{\psi} c_\text{s} \hat{k}_{\ast} \cdot \hat{n}}{(c_\text{s}
  + \hat{k}_{\ast} \cdot \hat{n}) \pi f_{\ast}} \sin (\pi f_{\ast}   L ( c_\text{s}^{- 1}  \hat{k}_{\ast} \cdot \hat{n}-1))~. \label{Rpfdm}
\end{eqnarray}
Differed from  $\mathcal{R}^{\text{(sdm)}}$, the amplitude of the trigonometric function in Eq.~(\ref{Rpfdm}) is also direction-dependent. In the limit of $c_\text{s} \rightarrow 0$, the phase of the trigonometric function in $\mathcal{R}^{\text{(pfdm)}}$ is dominated by the direction-dependent terms.

\begin{figure}
  \includegraphics[width=0.85\linewidth]{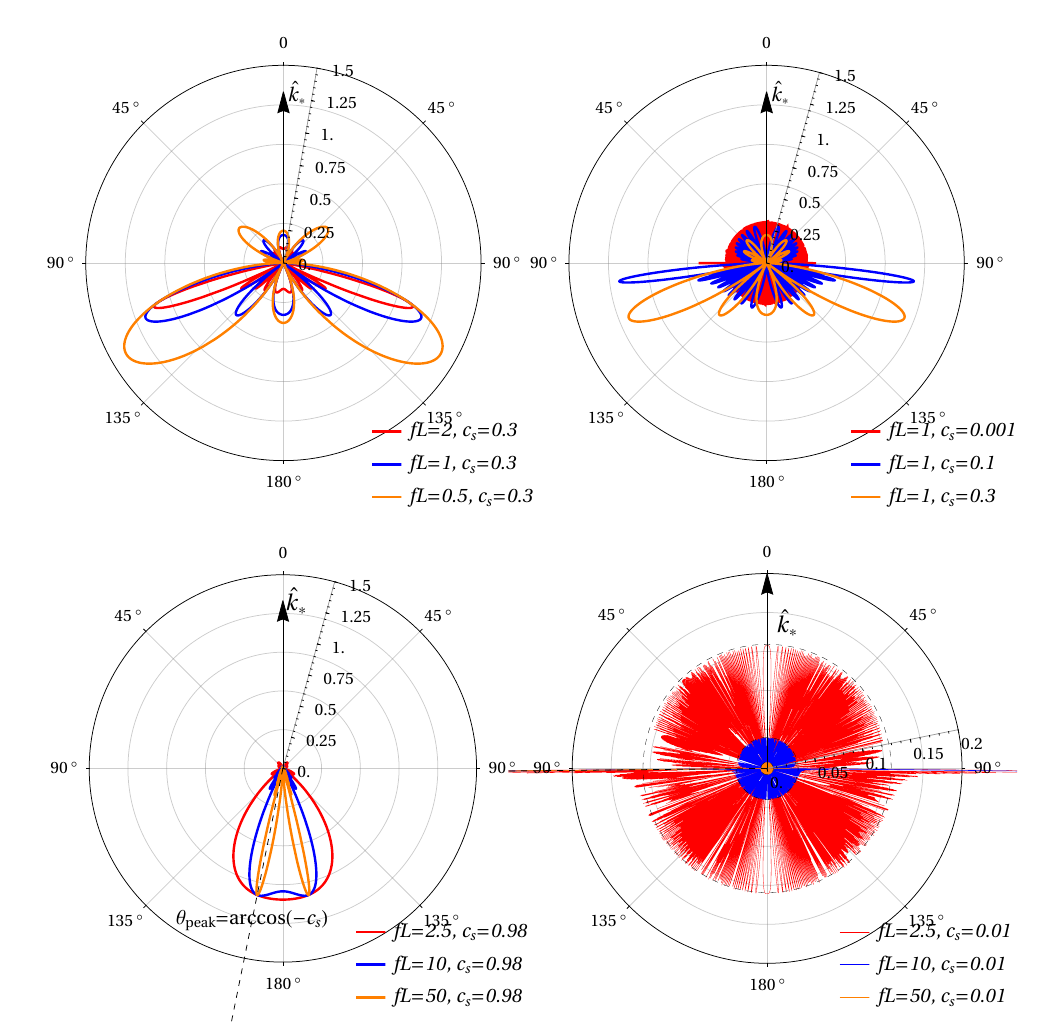}
  \caption{
    Curves of $| \mathcal{R}^\text{(pfdm)}|$ as a function of $\theta \equiv \arccos (\hat{k}_\ast \cdot \hat{n})$ in polar coordinates. For illustration, we set ${c_\text{s} \mathcal{A}_{\psi}L} = 1$ for the curves. In the left panels, $| \mathcal{R}^\text{(pfdm)}|$ is plotted for selected values of $fL$ with $c_\text{s}$ fixed. In the right panels, $| \mathcal{R}^\text{(pfdm)}|$ is shown for selected values of $c_\text{s}$ with $fL$ fixed.  
   \label{F6}} 
\end{figure}
In Figure~\ref{F6}, we show the directional dependence of the timing residual amplitude originating from oscillating perfect fluid dark matter. Both the sound speed $c_{\rm s}$ and $fL$ can affect the degree of directional dependence. Specifically, decrease in $fL$ or increase in $c_\text{s}$ results in more complicated features in the timing residual amplitude curves. The timing residual amplitude is enhanced as $\hat{k}_\ast \cdot \hat{n} + c_{\rm s} \rightarrow 0$, as shown in the bottom-right panel of Figure~\ref{F6}. In practical cases where $2fL$ is non-integer, setting $\hat{k}_\ast \cdot \hat{n} + c_{\rm s} \rightarrow 0$ would lead to a divergent timing residual amplitude. 

\subsubsection{Angular correlation from stochastic sources \label{IIIB2}}

To statistically study the $z$ originating from the oscillating perfect fluid dark matter in our galaxy, one can introduce the spatial correlation of $\bar{\psi}_{\textbf{k}}$ in the form of
\begin{eqnarray}
  \left\langle \bar{\psi}_{\textbf{k}} \bar{\psi}_{\textbf{k}'} \right\rangle
  & = & (2 \pi)^3 \delta \left( \textbf{k} + \textbf{k}' \right) P_{\psi} (k)~, \label{cor:psi}
\end{eqnarray}
where $P_\phi(k)$ is the power spectrum of the curvature perturbations, and it is also assumed to be Gaussian, stationary and isotropic, similar to Eq.~(\ref{cor:varphi}). By making use of Eq.~(\ref{zpfdm}) and (\ref{cor:psi})
we can derive the correlation of the $z$ to be
\begin{eqnarray}
  \langle z_\text{A} z_\text{B}^{\ast} \rangle & = & \frac{1}{2} \int_{-
  \infty}^{\infty}  {\rm d} f   S (f) \Gamma^\text{(pfdm)} (f, \theta_\text{AB})~,
\end{eqnarray}
where the ORF is
\begin{eqnarray}
  \Gamma^\text{(pfdm)} (f, \theta_\text{AB}) & = & \int  \frac{{\rm d} \Omega}{4\pi} \Bigg\{
  \frac{(\hat{k} \cdot \hat{n}_\text{A}) (\hat{k} \cdot \hat{n}_\text{B})}{\left( \frac{c_\text{s}
  f}{| f |} + \hat{k} \cdot \hat{n}_\text{A} \right) \left( \frac{c_\text{s} f}{| f |} +
  \hat{k} \cdot \hat{n}_\text{B} \right)} \nonumber \\ && \times\left( 1 - e^{2 \pi i   f   L_\text{A}
  \left( 1 + \frac{| f |}{c_\text{s} f} \hat{k} \cdot \hat{n}_\text{A} \right)} \right)
  \left( 1 - e^{- 2 \pi i   f   L_\text{B} \left( 1 + \frac{| f |}{c_\text{s} f
   } \hat{k} \cdot \hat{n}_\text{B} \right)} \right) \Bigg\}~, \label{ORFpfdm}
\end{eqnarray}
the spectral density of curvature perturbations can be given by
\begin{eqnarray}
  S (f) & = & \frac{2 c_\text{s}^2}{| f |} \mathcal{P}_{\psi} \left( \frac{2 \pi | f
  |}{c_\text{s}} \right)~, \label{SDpfdm}
\end{eqnarray}
and $\mathcal{P}_{\psi} (k) [ = {k^3P_{\psi} (k)}/{(2 \pi^2)}]$
is dimensionless power spectrum. Similar to the Hellings-Downs curve \cite{Hellings:1983fr}, the ORF in Eq.~(\ref{ORFpfdm}) is independent of the spectrum $\mathcal{P}_\psi(k)$.

\begin{figure}
  \includegraphics[width=1\linewidth]{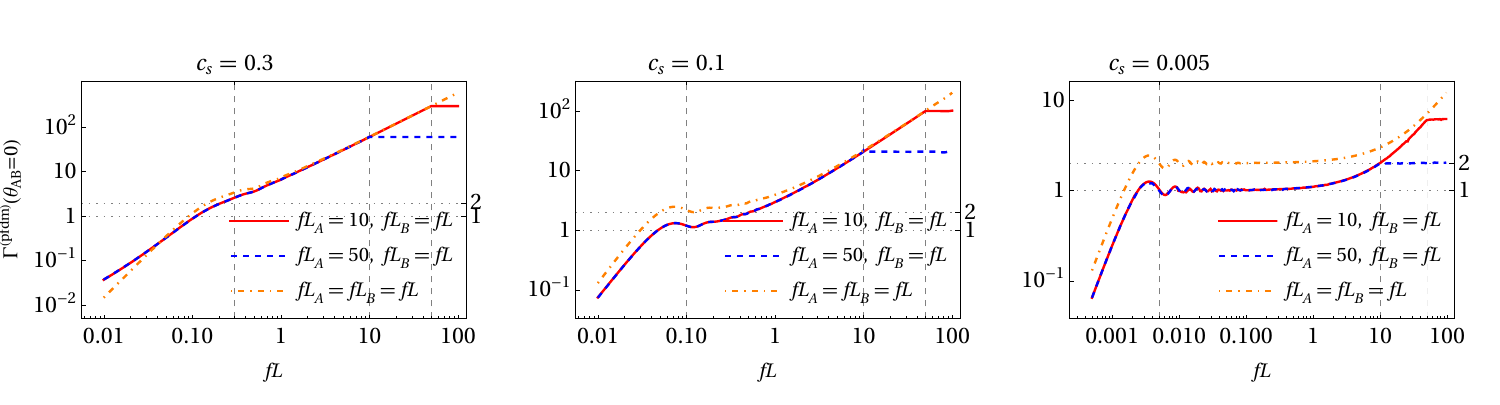}   
  \caption{ORFs at $\theta_\text{AB}=0$ as function of $fL$ for a fixed $c_\text{s}=$0.3(left panel), 0.1(medium panel) or 0.005(right panel). The gray dashed lines are formulated by $fL=c_\text{s}$ and $fL=fL_\text{A}$, respectively.
    \label{F7}} 
\end{figure}
\begin{figure}
  \includegraphics[width=0.85\linewidth]{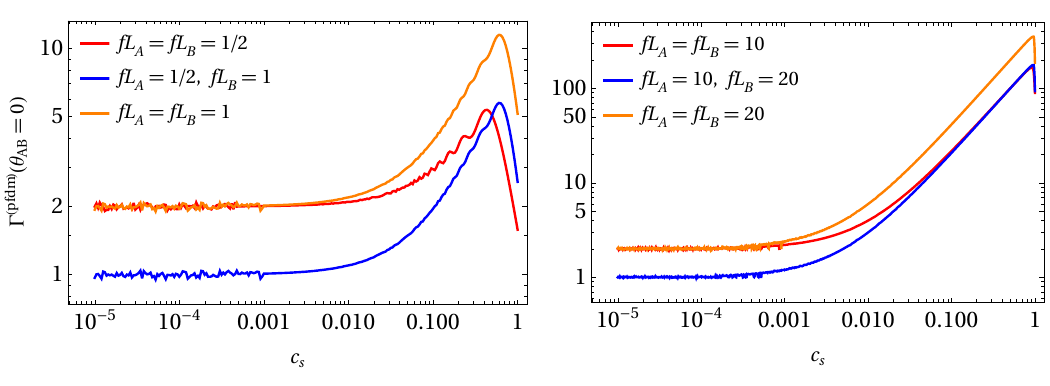}
  \caption{ORFs originating from perfect fluid dark matter at $\theta_\text{AB}=0$ as function of $c_\text{s}$ for a fixed $fL$.
    \label{F8}} 
\end{figure}
\begin{figure}
  \includegraphics[width=0.85\linewidth]{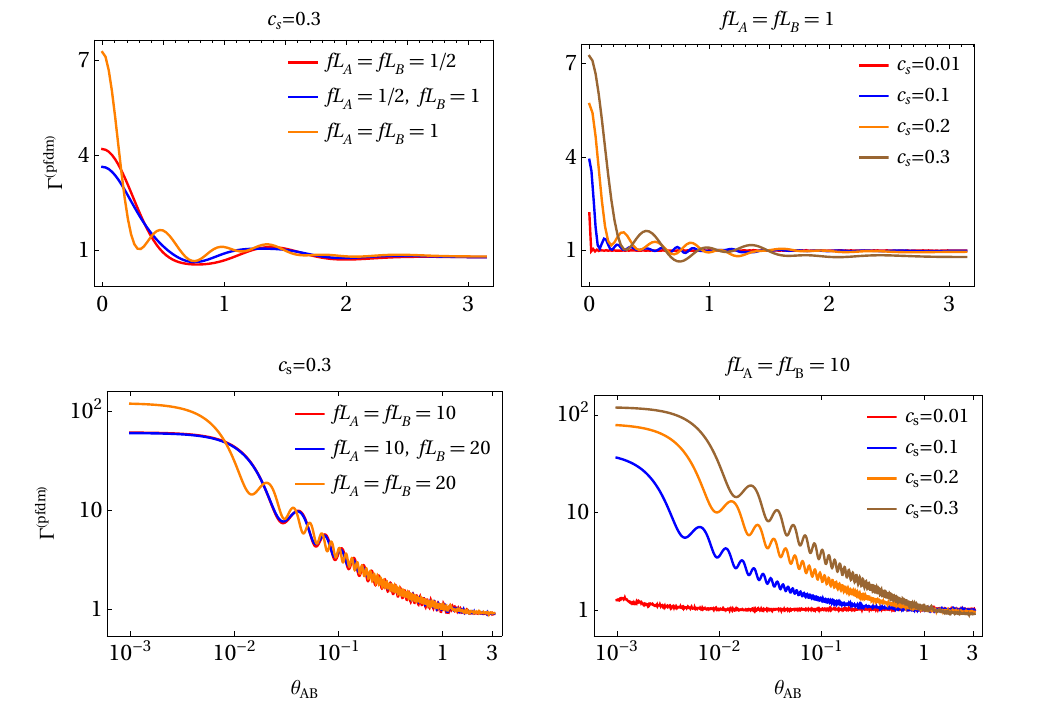}
  \caption{Angular correlation originating from perfect fluid dark matter formulated by the $\Gamma^\text{(pfdm)}$  for selected $fL_\text{A}$, $fL_\text{B}$ and $c_\text{s}$.  
    \label{F9}}
\end{figure}
Figure~\ref{F7} shows the ORFs originating from perfect fluid dark matter at $\theta_\text{AB}=0$ for selected distances of pulsar pairs. It is observed that $\Gamma^\text{(pfdm)}|_{\theta_\text{AB}=0}$ increases with $fL_\text{min} [\equiv \min(fL_\text{A}, fL_\text{B})]$ when $fL_\text{min} \lesssim c_\text{s}$. However, this effect is not important because $fL < c_\text{s}$ is not practical in PTA observations. As shown in the right panel of Figure~\ref{F7}, $\Gamma^\text{(pfdm)}|_{\theta_\text{AB}=0}$ becomes insensitive to variations in $fL_\text{min}$ when $c_\text{s}$ is small and $fL \gtrsim c_\text{s}$. Figure~\ref{F8} illustrates how $\Gamma^\text{(pfdm)}|_{\theta_\text{AB}=0}$ varies with the sound speed $c_\text{s}$. It additionally shows the decrease in $\Gamma^\text{(pfdm)}|_{\theta_\text{AB}=0}$ as $c_\text{s}$ approaches 1.
In Figure~\ref{F9}, we also present the angular correlation in PTAs. The correlation tends to approach 1 for $\theta_\text{AB} \neq 0$ and is enhanced only when the pulsar pair is very close to each other. Interestingly, similar results were also found in the studies on longitude modes of non-canonical gravitational waves \cite{2008ApJ...685.1304L,Chamberlin:2011ev} and Hellings-Downs curves in an expanding universe \cite{Zhu:2022wbd}. 


\section{Scalar dark matter vs. perfect fluid dark matter in PTA observations \label{IV}}

In this section, we will compare the scalar dark matter and the perfect fluid dark matter in PTA observations. Firstly, we need to clarify physical values of the parameters $\alpha$, $c_\text{s}$, and $fL$.
Because the dark matter velocity in our galaxy is in the order of $\mathcal{O}(10^{-3})$ \cite{Khmelnitsky:2013lxt,NANOGrav:2023hvm}, we set $2\alpha$ and $c_\text{s}$ to $10^{-3}$ in this context. The NANOGrav frequency band is [$2 \times 10^{-9}$, $5 \times 10^{-8}$] Hz \cite{NANOGrav:2023hvm,NANOGrav:2023gor}, and the distances of millisecond pulsars to the earth can be found in the ATNF Pulsar Catalogue \cite{Manchester:2004bp}. For the 68 millisecond pulsars used by NANOGrav \cite{NANOGrav:2023hde}\footnote{66 out of the 68 millisecond pulsars are listed in the ATNF Pulsar Catalogue.}, the distances range from [$0.157$, $7.158$] kpc. There are five pulsar pairs with the distance differences of less than $0.001$ kpc. Based on these facts, for our study, $fL$ ranges from [$32$, $3.6 \times 10^4$], with a minimum $f \delta L \simeq 0.2$ for a pulsar pair. Considering the indications of a monopolar signal near 4 nHz reported by NANOGrav \cite{NANOGrav:2023gor}, the range of $fL$ could reduce to [$65$, $2946$] with a fixed $f = 4 \text{ nHz}$.  
 
\begin{figure}
  \includegraphics[width=0.9\linewidth]{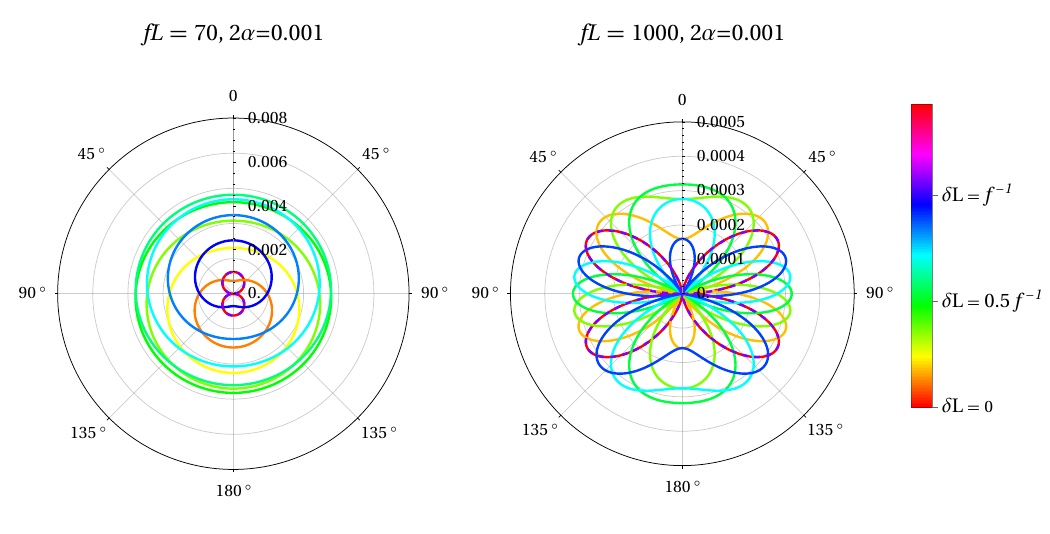}
  \caption{
    Directional dependence of the timing residual amplitude $|\mathcal{R}^\text{(sdm)}|$ with $2\alpha = 0.001$ for selected values of $f \delta L$ and $f L$ within the range [$65$, $2946$]. For illustration, we set $\kappa \mathcal{A}^2_\varphi L / 2 = 1$.  
    \label{F10}}
\end{figure}
\begin{figure} 
  \includegraphics[width=0.85\linewidth]{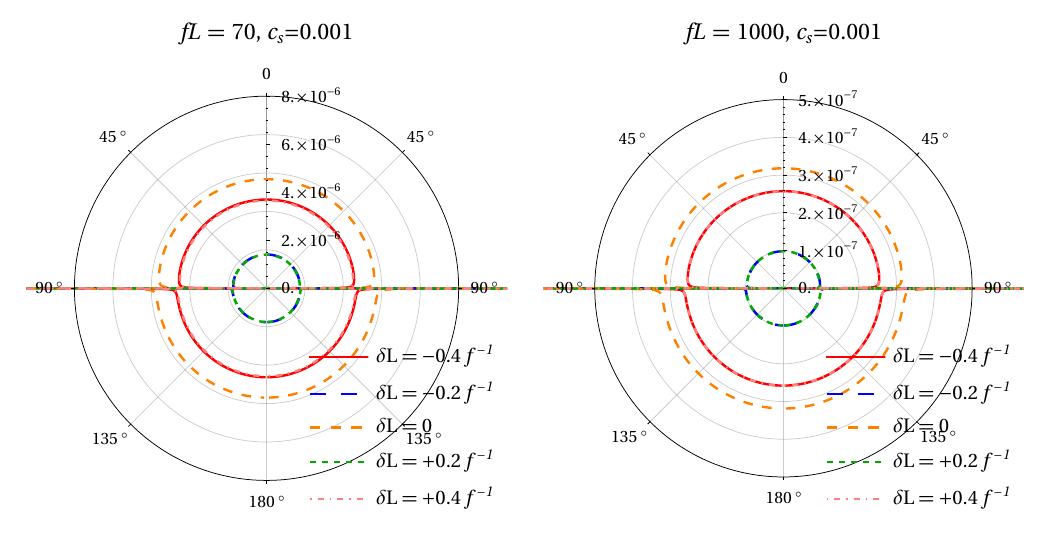}
  \caption{Directional dependence of the envelope curves of the timing residual amplitude $|\mathcal{R}^\text{(pfdm)}|$ with $c_\text{s} = 0.001$ for selected values of $f \delta L$ and $f L$ within the range [$65$, $2946$]. For illustration, we set $\mathcal{A}_\psi L = 1$.
    \label{F11}}
\end{figure}
\begin{figure}  
  \includegraphics[width=0.85\linewidth]{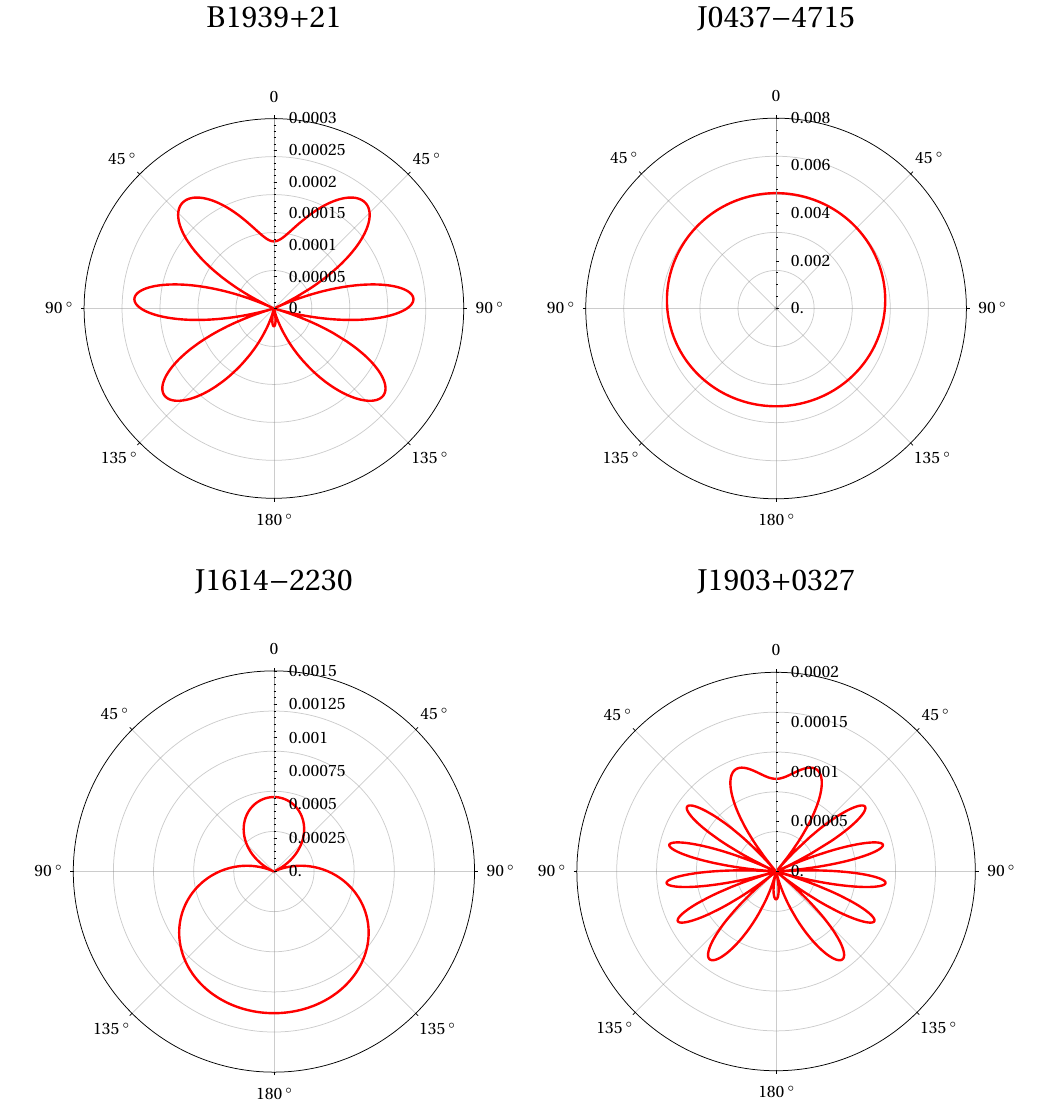} 
  \caption{Directional dependence of timing residual amplitudes $|\mathcal{R}^\text{(sdm)}|$ with given $2\alpha=0.001$ and $f=4 \text{ nHz}$. The distances $L$ are given by the known millisecond pulsars listed in ATNF Pulsar Catalogue \cite{Manchester:2004bp}.
    \label{F12}}
\end{figure}

We present the directional dependence of the timing residual amplitudes with practical parameters in Figures~\ref{F10} and \ref{F11}. The timing residuals originating from scalar dark matter $\mathcal{R}^\text{(sdm)}$ are sensitive to small variations in the distance $\delta L$, while those from perfect fluid dark matter $\mathcal{R}^\text{(pfdm)}$ are not. In other words, the $\mathcal{R}^\text{(sdm)}$ depends on the configuration of pulsars, whereas the $\mathcal{R}^\text{(pfdm)}$ can give a universal result. For illustration, we also present the directional dependence of $\mathcal{R}^\text{(sdm)}$ for four representative pulsars in Figure~\ref{F12}, where distinctive features are shown, obviously. A large value of $fL$ can enhance the degree of directional dependence, as exemplified by J0437-4715, the closest pulsar to the earth that has been observed. In a word, the effect from the subleading-order correction of $\mathcal{O}(k/m)$ is non-negligible in PTA observations. Assuming the same amount of dark matter comprised by scalar dark matter or perfect fluid dark matter, i.e., $\rho_\text{DM} \propto \mathcal{A}_\psi^\text{(pfdm)} \simeq \mathcal{A}_\psi^\text{(sdm)} \approx \kappa \mathcal{A}_\varphi^2 / 2$, we find $|\mathcal{R}^\text{(pfdm)}| / |\mathcal{R}^\text{(sdm)}| \simeq \mathcal{O}(c_\text{s})$ for $\hat{k} \cdot \hat{n} \neq -c_\text{s}$. It indicates that scalar dark matter is more sensitive in PTA observations than perfect fluid dark matter.

\begin{figure}
  \includegraphics[width=1\linewidth]{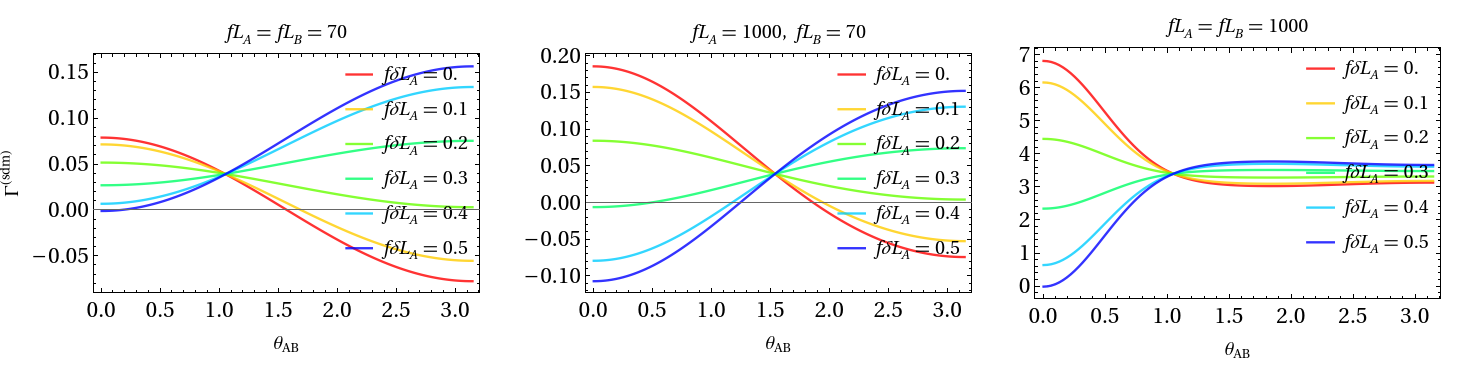}
  \caption{Angular correlation originating from scalar dark matter $\Gamma^\text{(sdm)}$  for practical parameters $fL_\text{A}$, $fL_\text{B}$ and $f\delta L_\text{A}$ with fixed $2\alpha=0.001$. 
    \label{F13}}
\end{figure}

\begin{figure}
  \includegraphics[width=0.85\linewidth]{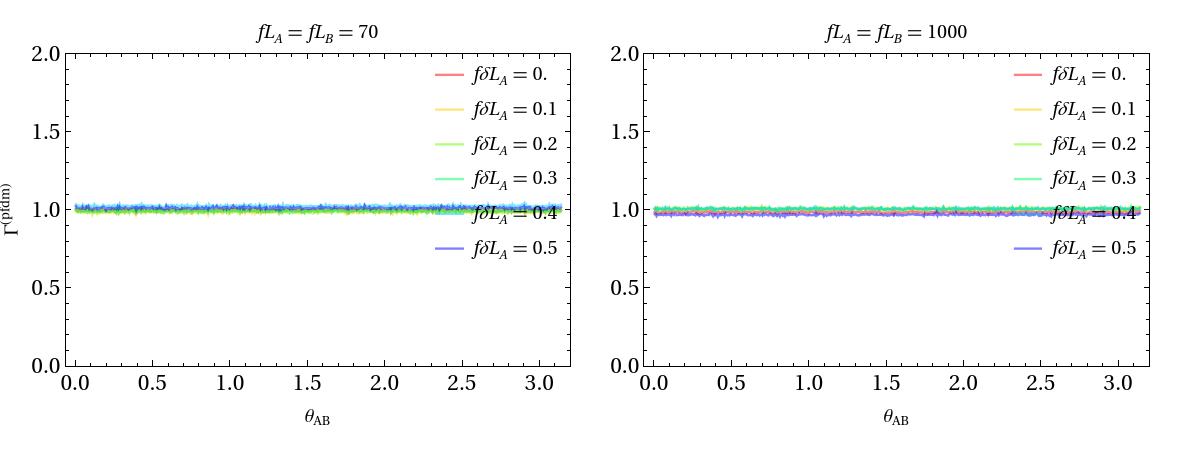}
  \caption{Angular correlation originating from perfect fluid dark matter $\Gamma^\text{(pfdm)}$  for practical parameters $fL_\text{A}$, $fL_\text{B}$ and $f\delta L_\text{A}$ with fixed $c_\text{s}=0.001$. 
    \label{F14}}
\end{figure}

Similarly, the angular correlation $\Gamma^\text{(sdm)}$ is also more sensitive to small variations in the pulsar's distance than the $\Gamma^\text{(pfdm)}$, as shown in Figures~\ref{F13} and \ref{F14}. According to the results in Section~\ref{IIIB2}, there are additional enhancements of $\Gamma^\text{(pfdm)}$ as $\theta_\text{AB} \to 0$. However, this enhancement seems to be of limited importance for current PTA observations \cite{NANOGrav:2023gor}. In this sense, perfect fluid dark matter might be a more suitable physical origin for monopolar signals in PTAs compared to scalar dark matter. 

\section{Conclusions and discussions \label{V}}

This study investigated the fluctuations in pulsar timing originating from the coherent oscillation of scalar dark with the subleading-order correction of $\mathcal{O}(k/m)$, as well as from the acoustic oscillation of non-adiabatic perfect fluid dark matter, based on the pure gravitational effect. Distinctive behaviors in pulsar timing residuals and angular correlations were presented for these two types of dark matter. We further compared these observables in the PTA frequency band and considering the known distances of pulsars.
For scalar dark matter, both the timing residual amplitude and the angular correlation are sensitive to small variations in the distance of pulsars, $\delta L$, due to the subleading-order correction. In contrast, for perfect fluid dark matter, it is insensitive to the $\delta L$. It is expected that the distinctive behaviors could be used to identify different types of dark matter in the future. 

We considered the dark matter in our galaxy as either the deterministic sources or stochastic sources. In PTA observations, the determined sources can be detected with timing residuals from a single pulsar \cite{Detweiler:1979wn}, while the stochastic sources could be detected using angular correlation of signals from a set of pulsar pairs \cite{Hellings:1983fr}. They are two different ways to analyze potential signals in PTAs. Both of them were also considered in PTA observations \cite{NANOGrav:2023hvm}.

We showed that the non-adiabatic perfect fluid dark matter can have impact on the pulsar timing. If it is of a deterministic source, there is an enhancement in the timing residual amplitude when $\hat{n} \cdot \hat{k} = -c_\text{s}$. Namely, the direction of the radio beam from a pulsar $\hat{n}$ is orthogonal to the propagation direction of the oscillating perfect fluid dark matter $\hat{k}$. On the other hand, if it is of stochastic source, its angular correlation in PTAs tends to be a constant, and is enhanced only when the pulsar pair is very close to each other. In this sense, perfect fluid dark matter might be a more suitable physical origin for monopolar signals in angular correlations compared to the scalar dark matter.

We studied the fluctuations in pulsar timing originating from the pure gravitational effect induced by dark matter. The leading-order effects in the weak-field expansion that could impact pulsar timing were considered. Here, the $k/m$-expansion for scalar field is also included within this leading-order weak-field effect. The equations of motion for scalar dark matter shown in Eq.~(\ref{moesdm}), differ from those for perfect fluid dark matter in Eq.~(\ref{moepf}). 
Specifically, there is non-vanishing anisotropic stress in the energy-momentum tensor for scalar dark matter, whereas perfect fluids are known in absence of the anisotropic stress \cite{Chang:2020tji}. Recent work has attributed the stochastic scalar dark matter to the fluctuations in the density $\delta \rho$ \cite{Kim:2023kyy}. This scenario seems to be correct in the limit $k/m \ll 1$. However, the $\delta \rho$ alone can not describe the exact dynamics of scalar dark matter, if considering the subleading-order correction of $\mathcal{O}(k/m)$ as shown in Eqs.~(\ref{metsdm0}), $\psi \neq - \phi$. In this case, a more rigorous derivation as presented in Section~\ref{II} should be employed \cite{Mukhanov:1990me,Malik:2008im}.

In our scenario, we considered the Fourier mode of curvature perturbation $\bar{\psi}_\textbf{k}$ as a  stochastic variable for perfect fluid dark matter, while treating the Fourier mode of the scalar field $\bar{\varphi}_\textbf{k}$ itself as a  stochastic variable. A key similarity between the two scenarios is that both $\psi$ (for perfect fluid dark matter) and $\varphi$ (for scalar dark matter) satisfy wave equations. The solutions of the wave equations, expressed in terms of plane-wave expansions, ensure that the spatial Fourier modes $\bar{\psi}_\textbf{k}$ and $\bar{\varphi}_\textbf{k}$ are time-independent. Only by adopting this scenario, can our approach be consistent with the assumption that the stochastic variables are stationary, as described by Eqs.~(\ref{cor:varphi}) and (\ref{cor:psi}).



The non-adiabatic perfect fluid can serve as an effective fluid description for axion-like particles in the WKB regime \cite{Hwang:2009js,Park:2012ru,Marsh:2015xka,Passaglia:2022bcr} or for multiple-component fluids \cite{Koshelev:2010wz,Maggiore:2018sht}, and it can also be constructed using effective field theory \cite{Kopp:2016mhm,Salehian:2020bon}. Therefore, it seems well-motivated to explore the potential signals from non-adiabatic perfect fluids in PTAs.

For scalar dark matter, the ORFs depend on the shape of the power spectrum $P_\varphi(k)$, which can be related to the velocity distribution of dark matter in our galaxy. Therefore, PTAs can serve as a probe for detecting the dark matter velocity distribution in our galaxy \cite{Kim:2023kyy}. In this study, we additionally found that the angular correlation of scalar dark matter can be influenced by the shape of the power spectrum $P_\varphi(k)$, even if it is isotropic. It is expected that future studies might have capability to explore shape or anisotropy of velocity distribution of dark matter with the angular correlation in PTAs.

\

{\it Acknowledgments. }
The author thanks Prof.~Qing-Guo Huang for useful discussions in the early stages of this work, and thanks Prof.~Zu-Cheng Chen and Dr.~Yu-Mei Wu for useful comments.
This work is supported by the National Natural Science Foundation of China under grants No.~12305073 and No.~12347101. 

\appendix

\section{Some calculation details: perfect fluid dark matter \label{App1}}
 
In this part, we will present the calculation details using perfect fluid dark matter as an example. Because there is a general approach to evaluating the perturbed Einstein equations and perturbed geodesic equations in Section~\ref{II} (see a pedagogical derivation \cite{Chang:2020tji,Zhu:2022bwf}), we will focus the calculation details about the $z$ in PTA observations.

\subsection{Expressing plane-wave expansion in terms of frequency and propagation direction}
The plane-wave expansion solution of the wave equation can be given by
\begin{eqnarray}
  \Psi & = & \int \frac{\textrm{d}^3 k}{(2 \pi)^3} \left\{ \Psi_{\textbf{k}} e^{-
  i   \left( c_s k   t - \textbf{k} \cdot \textbf{x} \right)} +
  \Psi_{- \textbf{k}} e^{i \left( c_s k   t - \textbf{k} \cdot
  \textbf{x} \right)} \right\}\nonumber\\
  & = & \int \frac{\textrm{d}^3 k}{(2 \pi)^3} \left\{ 2 \cos (c_s k   t)
  {\Psi}_{\textbf{k}} e^{i   \textbf{k} \cdot \textbf{x}} \right\}~, \label{A1}
\end{eqnarray}
where $\textbf{k}$ is wavenumber, and we have an explicit dispersion relation, $w=c_\text{s}k$. By making use of temporal Fourier transformation, we have 
\begin{eqnarray}
  \mathcal{F} [\Psi] (w) & = & \int \textrm{d} t   e^{i   w  
  t} \int \frac{\textrm{d}^3 k}{(2 \pi)^3} \left\{ 2 \cos (c_s k   t)
  \Psi_{\textbf{k}} e^{i   \textbf{k} \cdot \textbf{x}} \right\}\nonumber\\
  & = & \int \frac{\textrm{d}^3 k}{(2 \pi)^3} \left\{ 2 \pi \delta (| w | - c_s
  k) \Psi_{\textbf{k}} e^{i \textbf{k} \cdot \textbf{x}} \right\} \nonumber\\
  & = & \int \textrm{d} \Omega \left\{ \frac{1}{(2 \pi)^2} \left( \frac{w}{c_s}
  \right)^2 \Psi_{\textbf{k}} e^{i   c_s^{- 1} | w | \hat{k} \cdot
  \textbf{x}} \right\}~,
\end{eqnarray}
where $\textrm{d}\Omega$ is the volume element of angular coordinates with respect to $\hat{k}$, and we have integrated over the norm $k$. And then, we utilize the inverse Fourier transformation, namely,
\begin{eqnarray}
  \Psi \left( t, \textbf{x} \right) & = & \mathcal{F}^{- 1} [\mathcal{F}
  [\Psi] (w)] (t) \nonumber\\
  & = & \int_{- \infty}^{\infty} \frac{\textrm{d} w}{2 \pi} e^{- i   w
    t} \int \textrm{d} \Omega \left\{ \frac{1}{(2 \pi)^2} \left(
  \frac{w}{c_s} \right)^2 \Psi_{\textbf{k}} e^{i   c_s^{- 1} | w |
  \hat{k} \cdot \textbf{x}} \right\} \nonumber\\
  & = & \int_{- \infty}^{\infty} \textrm{d} f \int \textrm{d} \Omega \left\{ \left(
  \frac{f}{c_s} \right)^2 \left. \Psi_{\textbf{k}} \right|_{k = 2 \pi | f | /
  c_s} e^{- 2 \pi i   f \left( t - \frac{| f |}{c_s f} \hat{k} \cdot
  \textbf{x} \right)} \right\}~. \label{A2}
\end{eqnarray}
In the last equal sign, we have introduced the frequency $f=w/(2\pi)$.
Thus, the quantity $\Psi$ is expressed in terms the expansion of frequency $f$ and direction $\hat{k}$. Adopting symbolic convention used for stochastic GW \cite{Maggiore:2018sht}, one can introduce $\Psi(f,\hat{k})\equiv (f/c_\text{s})^2\Psi_\textbf{k}|_{k=2\pi|f|/c_\text{s}}$. Because the factor $(f/c_\text{s})^2$ is directionally independent, the transformation from Eq.~(\ref{A1}) to (\ref{A2}) does not affect the results of angular correlation.

\subsection{Evaluation of $z$ with deterministic source}
For deterministic source, one can utilize the ansatz as follows,
\begin{eqnarray}
  \bar{\psi}_{\textbf{k}} & \simeq & (2 \pi)^3 \mathcal{A}_{\psi} \delta
  \left( \textbf{k} - \textbf{k}_{\ast} \right) = \frac{1}{k^2} (2 \pi)^3
  \mathcal{A}_{\psi} \delta (k - k_{\ast}) \delta (\hat{k} - \hat{k}_{\ast})~.\label{A4}
\end{eqnarray}
Substituting the Eq.~(\ref{A4}) into Eq.~(\ref{zpfdm}), we obtain
\begin{eqnarray}
  z & = & \int_{- \infty}^{\infty} \textrm{d} f \int \textrm{d} \Omega \Bigg\{ \left(
  \frac{f}{c_s} \right)^2 (2 \pi)^3 \mathcal{A}_{\psi} \left( \frac{1}{\frac{2
  \pi | f |}{c_s}} \right)^2 \delta \left( \frac{2 \pi | f |}{c_s} - k_{\ast}
  \right) \delta (\hat{k} - \hat{k}_{\ast})  \nonumber\\
  &  & \left( - \frac{\hat{k} \cdot \hat{n}}{\frac{c_s f}{| f |} + \hat{k}
  \cdot \hat{n}} \right) e^{- 2 \pi i   f   t} \left( 1 - e^{2
  \pi i   f   L \left( 1 + \frac{| f |}{c_s f} \hat{k} \cdot
  \hat{n} \right)} \right) \Bigg\}\nonumber \\
  & = & - \frac{c_s \mathcal{A}_{\psi} \hat{k} \cdot \hat{n}}{c_s +
  \hat{k}_{\ast} \cdot \hat{n}} e^{- 2 \pi i   f_{\ast}   t} (1
  - e^{2 \pi i   f_{\ast}   L (1 + c_s^{- 1} \hat{k}_{\ast}
  \cdot \hat{n})}) + \left( \text{negative frequency term} \right)~.\label{A5}
\end{eqnarray}
One can further obtain Eq.~(\ref{zpfdm2}) by considering the real and positive frequency part of Eq.~(\ref{A5}).

\subsection{Evaluation of $z$ with stochastic source}

For stochastic source, we expect a statistical description of the dark matter. Here, we adopt the spatial correlation of the curvature perturbation, namely,
\begin{eqnarray}
  \left\langle \bar{\psi}_{\textbf{k}} \bar{\psi}_{\textbf{k}}^{\ast}
  \right\rangle & = & (2 \pi)^3 \delta \left( \textbf{k} - \textbf{k} \right)
  P_{\psi} (k) \nonumber \\
  & = & \frac{1}{| k |^2} (2 \pi)^3 \delta (k - k') \delta (\hat{k} -
  \hat{k}') P_{\psi} (k) \nonumber\\
  & = & \left( \frac{c_s}{| f |} \right)^4 \delta (| f | - | f' |) \delta
  (\hat{k} - \hat{k}') \frac{1}{4 \pi} \frac{c_s^2}{| f |} \mathcal{P}_{\psi}
  \left( \frac{2 \pi | f |}{c_s} \right)~. \label{A6}
\end{eqnarray}
The correlation is the originally defined in wavenumber $\textbf{k}$. One can also rewrite it in terms of $f$ and $\hat{k}$ by making use of the dispersion relations.  We use the ensemble-average approach to study the stochastic sources \cite{Allen:2022dzg}. One should forget the ansatz in Eq.~(\ref{A4}), as it no longer holds in this approach. 

By making use of Eq.~(\ref{A6}), one can derive the correlation of the $z$ for a pulsar pair, namely,
\begin{eqnarray}
  \langle z_A z_B^{\ast} \rangle 
  & = & \int \textrm{d} f \int \textrm{d} f' \int \textrm{d} \Omega \int \textrm{d} \Omega'
  \Bigg\{ \frac{f^2 {f'}^2}{c_s^4} \left\langle \bar{\psi}_{\textbf{k}}
  \bar{\psi}_{\textbf{k}}^{\ast} \right\rangle \left( \frac{\hat{k} \cdot
  \hat{n}_A}{\frac{c_s f}{| f |} + \hat{k} \cdot \hat{n}_A} \right) \left(
  \frac{\hat{k}' \cdot \hat{n}_B}{\frac{c_s f'}{| f' |} + \hat{k}' \cdot
  \hat{n}_B} \right)  \nonumber\\
  &  & \left( 1 - e^{2 \pi i   f   L_A \left( 1 + \frac{| f
  |}{c_s f} \hat{k} \cdot \hat{n}_A \right)} \right) \left( 1 - e^{- 2 \pi i
    f  ' L_B \left( 1 + \frac{| f' |}{c_s f'} \hat{k} \cdot
  \hat{n}_B \right)} \right) e^{- 2 \pi i (f - f') t} \Bigg\} \nonumber\\
  & = & \int \textrm{d} f \Bigg\{ \frac{1}{4 \pi} \frac{c_s^2}{| f |}
  \mathcal{P}_{\psi} \left( \frac{2 \pi | f |}{c_s} \right) \int \textrm{d} \Omega
  \Bigg\{ \left( \frac{\hat{k} \cdot \hat{n}_A}{\frac{c_s f}{| f |} + \hat{k}
  \cdot \hat{n}_B} \right) \left( \frac{\hat{k} \cdot \hat{n}_B}{\frac{c_s
  f}{| f |} + \hat{k} \cdot \hat{n}_B} \right) \nonumber\\
  &  & \left( 1 - e^{2 \pi i   f   L_A \left( 1 + \frac{| f
  |}{c_s f} \hat{k} \cdot \hat{n}_A \right)} \right) \left( 1 - e^{- 2 \pi i
    f   L_B \left( 1 + \frac{| f |}{c_s f} \hat{k} \cdot
  \hat{n}_B \right)} \right) + \underset{\text{average out}}{e^{4 \pi i
    f   t} (\ldots)} \Bigg\}\Bigg\}\nonumber\\
  & =: & \frac{1}{2} \int \textrm{d} f   S (f) \Gamma (f,
  \theta_{\textrm{AB}})~,
\end{eqnarray}
where the terms involving the factor $e^{4\pi i f t}$ could be averaged out over the time. Thus, the spectral density $S(f)$ and ORFs $\Gamma(f,\theta_\text{AB})$ can be obtained as shown in Eqs.~(\ref{ORFpfdm}) and (\ref{SDpfdm}).

One can also obtain a consistent definition of spectral density by making use of the correlation of $\psi (f, \hat{k})$, similar to Eq.~(23.28) in Ref.~\cite{Maggiore:2018sht}, namely,
\begin{eqnarray}
  \langle \psi (f, \hat{k}) \psi^{\ast} (f', \hat{k}') \rangle & = &
  \left( \frac{| f |}{c_s} \right)^2 \left( \frac{| f' |}{c_s} \right)^2
  \left\langle \bar{\psi}_{\textbf{k}} \bar{\psi}_{\textbf{k}}^{\ast}
  \right\rangle \nonumber\\
  & = & \delta (| f | - | f' |) \delta (\hat{k} - \hat{k}') \frac{1}{4 \pi}
  \frac{c_s^2}{| f |} \mathcal{P}_{\psi} \left( \frac{2 \pi | f |}{c_s}
  \right) \nonumber\\
  & = & \delta (| f | - | f' |) \frac{\delta (\hat{k} - \hat{k}')}{4 \pi}
  \frac{1}{2} S (f)~.
\end{eqnarray}

\bibliography{cite}

\end{document}